\documentclass[12pt,preprint]{aastex}

\newcommand{\kms}{km s$^{-1}$}
\def\arcs{\char'175\ }
\def\arcsc{\char'175 }
\def\arcm{\char'023\ }

\def\hub{\ifmmode H_\circ\else H$_\circ$\fi}
\def\kms{~km~s$^{-1}$\ }
\def\kmsn{~km~s$^{-1}$}
\newcommand{\noprint}[1]{}

\usepackage{lscape,graphicx,afterpage}
\begin{document}

\title{Star Formation in Partially Gas-Depleted Spiral Galaxies}  

\author{James A. Rose, Paul Robertson\altaffilmark{1}, Jesse Miner, \& Lorenza Levy}
\altaffiltext{1}{Now at Department of Astronomy, University of Texas at Austin}
\affil{University of North Carolina at Chapel Hill}
\affil{Department of Physics and Astronomy, CB 3255, Chapel Hill, NC
  27599}
\email{jim@physics.unc.edu,paul@astr.as.utexas.edu,jminer@physics.unc.edu,
lorenza.levy@yahoo.com}

\begin{abstract}  Broadband B and R and H$\alpha$ images have been obtained
with the 4.1-m SOAR telescope atop Cerro Pachon, Chile
for 29 spiral galaxies in the Pegasus I galaxy cluster and for 18 spirals
in non-cluster environments.  Pegasus I is a spiral-rich cluster
with a low density intracluster medium and a low galaxy velocity dispersion.
When combined with neutral hydrogen (HI) data obtained with the Arecibo 305-m 
radiotelescope, acquired by \citet{ll07} and by \citet{sp05a},
we study the star formation rates in disk galaxies as a function of their HI
deficiency.  To quantify HI deficiency, we use the usual logarithmic deficiency
parameter, $DEF$.  The specific star formation rate (SSFR) is quantified by the
logarithmic flux ratio of H$\alpha$ flux to R band flux, and thus roughly
characterizes the logarithmic SFR per unit stellar mass.  We find a clear
correlation between the global SFR per unit stellar mass and $DEF$, such that
the SFR is lower in more HI-deficient galaxies.  This correlation appears to
extend from the most gas-rich to the most gas-poor galaxies.  We also find a
correlation between the central SFR per unit mass relative to the global
values, in the sense that the more HI-deficient galaxies have a higher
{\it central} SFR per unit mass relative to their {\it global} SFR values than do gas-rich
galaxies.  In fact, approximately half of the HI-depleted galaxies have
highly elevated SSFRs in their central regions, indicative of a transient
evolutionary state.  In addition, we find a correlation between 
gas-depletion and the size of the H$\alpha$ disk (relative to the R band disk);
HI-poor galaxies have truncated disks.  Moreover, aside from the elevated
central SSFR in many gas-poor spirals, the SSFR is otherwise lower in the
H$\alpha$ disks of gas-poor galaxies than in gas-rich spirals.  Thus both
disk truncation {\it and} lowered SSFR levels within the star-forming part of
the disks (aside from the enhanced nuclear SSFR) correlate with HI deficiency,
and both phenomena are found to contribute equally to the global suppression
of star formation.
We compare our results found in the low richness Pegasus I cluster and in
non-cluster environments with SFRs found for HI-deficient spirals in the Virgo 
cluster by \citet{kk04a, kk04b, fg08}.

\end{abstract}

\keywords{galaxies: clusters: general, galaxies: evolution, galaxies:
  ISM, radio lines: galaxies}

\section{Introduction}

In the cores of nearby galaxy clusters early-type (E/S0) galaxies are the 
dominant population \citep[e.g.,][]{o74, d80, gom03, got03}. Furthermore, even the star-forming disk
galaxies tend to be depleted in atomic hydrogen (HI), typically by an order of
magnitude \citep[e.g.,][]{gh85, ga87, s01, ga06}.  In contrast, at modest redshifts (z$\gtrsim$0.3) the fraction of
blue, star-forming disk galaxies is substantially higher \citep{bo78, bo84, dg83, d99}.  This rapid recent
evolution in the disk galaxy population of clusters provides an excellent
opportunity to study the decline of star formation in galaxies, and its
causes, and has thus spurred numerous observational and theoretical studies.

The two most prevalent mechanisms for driving disk galaxy evolution in
clusters are (1) interaction between the galaxy interstellar medium (ISM) with
the diffuse hot intracluster medium (ICM) and (2) gravitational tidal effects.
In the first mechanism either ram pressure directly strips the galaxy ISM 
\citep{gg72}
via momentum transfer, thermal conduction, and/or viscous stripping
removes the ISM through energy transfer \citep[e.g.,][]{n82, cs77}.
In the second mechanism, gas removal is achieved either through major or minor
galaxy-galaxy encounters \citep[e.g.,][]{tt72, mh94, mh96},
through tidal interaction of the galaxy with the
cluster potential \citep{bv90} or through ``harrassment'' of a galaxy from numerous
impulsive encounters \citep{mo96, mo98}.  Several cases of asymmetric
HI, H$\alpha$, and radio continuum emission in cluster galaxies have provided
detailed and convincing evidence for ram pressure effects \citep{g95, k04, 
vo04, ck06, c07, yo08},
and gas dynamical simulations that reproduce the observations in detail
are becoming increasingly sophisticated \citep[][among others]{ab99, vo01, ss01, rh05, kr08a, kr08b}.
Nevertheless, tidal interactions are observed to be increasingly common in
clusters at higher redshift \citep[e.g.,][]{lh88, vd00}, and thus
appear to play an important role as well, especially when galaxies first
enter the main cluster in small groups with relatively low velocity dispersion
\citep[e.g.,][]{vg04, v04}.  It is in any case clear that ram pressure stripping is more effective
on extraplanar and diffuse interstellar gas in galaxies, and tidal interactions
indeed produce diffuse extraplanar gas \citep[e.g.,][]{sb04}.  Hence recent 
studies tend to advocate a synergy between ram pressure and tidal effects
in depleting the ISM in cluster disk galaxies \citep[e.g.,][]{c07, ka08}.
Furthermore, while in principle these mechanisms may completely deplete the ISM
of galaxies in the right circumstances, recent studies have tended to support
the proposal by \citet{ltc80} that ``starvation'' or ``strangulation''
of the disk occurs when the (more vulnerable) gas resupply from the outer disk 
and hot gaseous halo is cut off by the combined effects of ram pressure and
tidal interaction \citep{km08}.  Evidence for gas depletion and
altered SFR in cluster spirals, and for the mechanisms driving the observed
depletion, is extensively reviewed in \citet{bg06} and in \citet{vg04}.

While the fraction of S0 galaxies is highest in the cores of present epoch 
galaxy clusters, S0s do exist as well in poor clusters and in lower density 
environments.  Given that the balance between ram pressure and tidal effects
should vary with environment, an analysis of gas depletion characteristics 
for galaxies in low density versus rich cluster environments should provide
additional constraints on the gas depletion process.  The lower velocity
dispersions of galaxies in groups should favor tidal interaction over ram
pressure effects, when compared to the high velocity dispersion environments
of rich clusters.  Nevertheless, \citet{he06} has found that partial ram
pressure stripping of the less massive spirals should be possible in group
environments. In fact, several observational studies
have found both individual and statistical evidence for partially HI-depleted
galaxies in the group environment.  One notable example is the spiral galaxy
NGC 2276 in the NGC 2300 galaxy group, which has an asymmetric morphology and
high star formation rate
indicative of interaction between NGC 2276 and its environment \citep{d97},
is moderately HI deficient by a factor of $\sim$2 
\citep{rpm06}, and is located in the first loose
galaxy group for which extended X-ray emission has been discovered 
\citep{mu93, rpm06}.  Similarly, HI imaging of
spirals in the Holmberg 124 group \citep{ka05}, Eridanus group \citep{od05},
and four X-ray bright groups \citep{se07} find modest HI deficiencies, as well as peculiar HI
morphologies indicative of asymmetrical and/or extraplanar gas.  Since
ram pressure alone appears incapable of clearing out the HI disks to the level
observed, \citet{ka05} and \citet{sb06} suggest that a combination of
tidally assisted ram pressure stripping and evaporation by thermal conduction
may be the source of the observed HI deficiencies.  In addition, \citet{sb06}
find statistical evidence for a greater HI depletion
for galaxies inhabiting X-ray bright groups compared to groups without
detected X-ray emission.  Evidence has also been presented for HI-depleted
galaxies at such large radial distance in the Virgo cluster that the galaxies
cannot have yet experienced the central cluster environment \citep{so02}.
However, distance and modeling uncertainties may be sufficiently
large that the possibility of these galaxies having passed through the
Virgo cluster core remains open \citep{sa04, ckh08}.
Finally, \citet{sp05b} show that the HI mass
function in galaxies exhibits some environmental trends, even at local densities
that are substantially below the rich cluster level.

A question that naturally arises is whether
the HI deficiencies in galaxies translate as well into reduced star formation
rates, and if so, how the reduction of star formation is distributed within
the disk.  
As  Koopmann \& Kenney (2004a,b; hereafter KK04ab), Fumagalli \& Gavazzi
(2008; hereafter FG08), and others \citep[e.g.,][]{th08} have pointed out,
the characteristics of both global SFR
in HI-depleted galaxies and the detailed SFR profiles in those galaxies can
shed light on the mechanism(s) causing the loss
of gas in the high DEF spirals; HI imaging is similarly pertinent 
\citep{ca94, ba01}.  A particular emphasis of these previous
studies has been to characterize whether the reduced SFR that accompanies
HI depletion in Virgo cluster spirals is a result of disk truncation, or whether
a more general quenching is occurring of the SFR throughout a normal sized disk.
In fact, there is some disagreement between FG08 and KK04ab on this question.
KK04ab have carried out a comprehensive
investigation of star formation rates in Virgo cluster spirals on the basis
of H$\alpha$ and R band imaging. 
They find relatively few examples of ``quenched'' galaxies, i.e.,
with globally lowered SFR in a disk of normal extent.  Rather, approximately
50\% of the disks in their Virgo spirals are truncated in H$\alpha$.
Furthermore, KK04a find a general correspondence between H$\alpha$ and HI disk
sizes and between the morphologies of their H$\alpha$ disks and the HI disk
types found by \citet{ca94}.  On the other hand, while FG08 do find that
the most HI-depleted disks in their Virgo sample are truncated in H$\alpha$ as 
well, the more modestly depleted HI disks exhibit a more global depression of
SFR in an otherwise normal disk size.  FG08 thus argue that disk quenching
is the key process in Virgo cluster spirals, rather than disk truncation. 
Clearly, a resolution of the roles of disk truncation versus
quenching, in a broad range of environments in addition to the Virgo cluster, 
would be of considerable importance in assessing what mechanism is chiefly
responsible for the removal of atomic gas in galaxy disks.

In a recent study, \citet[][hereafter L07]{ll07} found evidence 
for mild (factor of 2)
HI deficiencies for many spirals in the Pegasus I cluster, which is a low
velocity dispersion poor cluster with low X-ray emission.  HI imaging of a few
of the gas-deficient spirals showed asymmetries and reduced HI-to-optical disk diameters
that further emphasized the anomalous properties of those galaxies.  The
low ram pressure in Pegasus I, as in the galaxy groups mentioned above, provides
an interesting intermediate environment in which to assess the sources of
HI deficiencies in disk galaxies.  

In this paper, we present optical R band and
H$\alpha$ imaging of many of the Pegasus I cluster spirals, as well as other 
spirals with a variety of HI deficiencies.   The goal is to evaluate the
connection between star formation properties (i.e., both global and radial star
formation profiles) and atomic gas depletion for galaxies in environments 
outside of rich clusters, with particular emphasis on the Pegasus I galaxy 
cluster that formed the basis of the L07 study.  Our objective is
to provide additional constraints on the gas depletion process, with the focus
on how such activity occurs in the group environment.
In \S 2 we describe the galaxy samples observed by us, the HI and optical 
observational data, and the data analysis methods.
In \S 3 we present both the radial star formation profiles and global rates
extracted from the H$\alpha$ and R band imaging, and relate them to the data
on HI deficiency.  Finally, in \S 4 a discussion of our results is given in
the context of other evidence for the sources of HI deficiency in spiral 
galaxies, and a brief summary is given in \S 5.

\section{Observations}

\subsection{Samples of Galaxies}

Our analysis of the connection between atomic gas depletion and SFRs in
galaxies is based on two samples of spiral galaxies.  The first is the
list of 54 spirals in the Pegasus I galaxy cluster, as well as 17 spirals
in a variety of environments approximately 1$^{hr}$ to the East of Pegasus,
that formed the basis of the L07 study.  Global HI measurements were
obtained for all of these galaxies with the
Arecibo 305~m radiotelescope for the L07 study.
In this paper we supplement those previous 21 cm observations with optical B 
and R band and H$\alpha$
imaging of 29 of the Pegasus I galaxies and 3 of the galaxies to the East of
Pegasus.  We refer to this sample of 32 galaxies as the L07 
sample.  As discussed in L07 the 
Pegasus I cluster can be divided into three distinct groups, a central group
with 3400$\leq$cz$\leq$4400 \kms, a foreground group, with 2500$\leq$cz$<$3400
\kms, and a background group, with 4400$<$cz$\leq$6000 \kmsn.  The background
group appears to link up with the Perseus-Pisces Supercluster \citep{hg86}.

A second sample of spiral galaxies, for which homogeneous global HI data is
available from the literature, was observed in B, R, and H$\alpha$ to fill out the
optical imaging program when the primary Pegasus I targets were at excessive
airmass. We refer to this second sample of 15 galaxies as the supplementary sample.
Specifically, we selected non-cluster Sa-Sc 
galaxies from the \citet[][hereafter S05a]{sp05a} catalog for which 
Arecibo observations have been made, and
covering a variety of HI deficiencies (The nature of the S05a
catalog is further discussed in \S2.2).  We also restricted the sample to the
redshift range 2700$<$cz$<$5500 \kms, which corresponds to the useful range
of the 100~\AA \ wide H$\alpha$ interference filter used for the observations.
Furthermore, we intentionally selected galaxies covering a range in neutral
hydrogen deficiency (HI deficiency is defined in the next Section).  Hence our
supplementary sample is far more biased towards galaxies with larger HI 
deficiency than a random selection from the S05a catalog would produce.

A key parameter for the galaxies in this
supplementary sample (as well as for the galaxies in the L07 sample)
is the local environment in which the galaxy is located.
While different techniques have
been applied to define local environment, including projected galaxy surface
density on the sky, we have chosen to use a three-dimensional number
density in which the position of each galaxy is calculated from its
sky position and recessional velocity.
To construct a density field, we used the Updated Zwicky
Catalog of \citet[hereafter UZC]{fa99}, which is
95\% complete to a limiting magnitude of $m_{Zw}$ = $15.5$ mag, includes
galaxies of all types, and covers most of the northern sky.
Using the sky positions and local-group--corrected redshifts of the
objects in the UZC, and adopting a value of the Hubble
constant of \hub=75 km s$^{-1}$, we assign each galaxy a 3D position in a
Cartesian coordinate system. To calculate the local density, we find the
mean distance to an object's six nearest neighbors, and use that distance to
define the radius of a sphere containing the ``local'' region. The number
of objects contained within the sphere is divided by the physical volume of
the sphere to obtain a number density in units of Mpc$^{-3}$.  This number
density is then corrected for the (in)completeness of the galaxy luminosity
function, relative to the degree of completeness at $cz=3000$\kms.  
Specifically, we have multiplied all of the calculated densities by a
LF correction factor which is the ratio of the integrated observable
LF (i.e., the number of galaxies per Mpc$^{3}$ brighter than the limiting
magnitude of the sample) at 3000 km s$^{-1}$ to the integrated LF at the
galaxy's redshift. We use a Schechter LF \citep{sc76} derived from the
UZC, with $\alpha = -1.0$ and $M^{*}_{B} = -18.8$ \citep{ma94}.

In summary, our dataset is based on two samples.  The first, L07, sample 
consists of Arecibo HI data and optical B, R, and H$\alpha$
imaging for 29 galaxies in the Pegasus I cluster and 3 non-cluster galaxies
to the east of Pegasus.  The second, supplementary, sample 
consists of 15 target of
opportunity non-cluster galaxies for which Arecibo data, compiled in a homogeneous
manner in the S05a catalog, also exists.  Our combined sample is distinct from those of KK04ab and FG08 
in that both of the latter samples are entirely based on the Virgo cluster
environment.  In contrast, our galaxies span a range of lower density 
environments, with greatest emphasis on the spiral-rich low velocity
dispersion Pegasus I cluster.  The range in luminosity of our galaxies is
quite similar to that in KK04ab, while the luminous Virgo spirals studied in
FG08 are typically 1-1.5 mag more luminous than our spirals.

Coordinates (equinox 2000.0) for our galaxy sample, as 
well as morphological types, redshift information, and the above-described
local number density are listed in 
Table~\ref{tab:one}.  The morphological T type, given in column (5), and the 
heliocentric redshift, in column (6), are taken from HyperLeda.  The Log Density
(in galaxies per Mpc$^{3}$) is given in column (7). A description 
of the HI and optical data is given below.

\subsection{HI Data}

Global HI measurements of 54 spiral galaxies in the Pegasus I cluster, as
well as for 17 spirals approximately 1$^{hr}$ east of Pegasus, were obtained
with the Arecibo 305~m telescope by L07.  Details on the
observations are given in that paper.  Here we concentrate on the HI data
for the 29 Pegasus I
spirals and 3 non-Pegasus spirals for which we have newly acquired broadband
optical and H$\alpha$ imaging.

Neutral hydrogen data for the 15 spirals in our second sample
are taken from the S05a catalog, which consists of global HI 
fluxes for over 8,000
spiral galaxies observed with several radio telescopes over a span of
some 20 years. The flux measurements have been re-processed 
to correct for instrumental effects such as beam attenuation and
pointing offsets, as well as HI self-absorption in the observed galaxies. Thus,
the corrected fluxes can be assumed to be homogeneous. While the catalog also
includes data from other radio telescopes, we have only used Arecibo
observations to avoid any instrument-to-instrument offsets.  

For both the L07 galaxies and the supplementary 15 galaxies from the
S05a catalog the global 21 cm flux is used to calculate the total HI 
mass of each galaxy.
Specifically, $M_{HI} = 2.36\times10^{5}(hr)^{2}F_{HI}M_{\sun}$,
where $r$ is the radial distance in kpc and $F_{HI}$ is the corrected 21
cm flux density in Jy km s$^{-1}$ \citep{so96}. The linear
optical diameter is given by $D_{0} = 0.291(hr)a$ kpc,
where $a$, the angular major axis diameter, is also provided by the
S05a catalog. 

\subsubsection{Determination of HI Deficiency}

The focus of this paper is to investigate the relationship between neutral hydrogen
depletion in galaxy disks and the global and local star formation rates.
To quantify the degree to which a particular galaxy is deficient in HI
we must compare the observed HI content with the amount of HI that is
{\it expected} in a galaxy of its morphological 
type and size. The logarithmic difference between the expected
and observed gas content will then represent the gas deficiency of the galaxy
(DEF).  Here we describe how the expected HI content is determined for the 
various galaxies in the L07 sample and in the supplementary sample of
galaxies extracted from the S05a catalog.

L07 calculate DEF values for the galaxies in the Pegasus
region using the prescription detailed in \citet{so96}, as follows.
First,  we recall that there is a logarithmic correlation between
a galaxy's HI mass and its linear diameter, with a slight dependence of the
slope of the fit on morphological type. If a fit of $\log(M_{HI})$ vs. 
$\log(D_{0})$ is performed on a sample of normal galaxies of the same
morphological type, then the best fit line is taken to represent the
expected HI mass as a function of linear diameter. At a given diameter, if the
measured HI mass is less than that predicted by the fit to the sample, then
the galaxy is defined to have a positive deficiency (DEF), 
given by the logarithmic
difference between the expected and the observed values:
\begin{equation}\label{def}
DEF = (\log{M_{HI}(D_{0},T)})_{exp} - \log{M_{HI}},
\end{equation}
Before one can perform these fits, however, a suitable sample of galaxies 
must be chosen to represent a ``normal'' population in terms of HI content.
Due to the increased likelihood of a galaxy to be gas deficient with 
increasing local density, this sample must be restricted to
low density environments to base the deficiency measurement on
galaxies whose gas content is dictated primarily by the fluctuations of
secular evolution, not environmental effects. \citet{so96} use
a survey of 4620 galaxies from the Perseus-Pisces supercluster region, and
define a low density sample of 2582 galaxies based on the sky-projected surface density
of galaxies. All of the galaxies in low surface density regions are combined
according to morphological type and used to perform the 
$\log(M_{HI})$-$\log(D_{0})$ fit.  A large galaxy sample is necessary due to
the existence of a natural scatter around the mean $\log(M_{HI})$-$\log(D_{0})$ 
relation.  The slopes and zero-points from these fits
are then used to determine the expected HI content for all of the galaxies
in the sample. The DEF values of L07 come directly from these
results.

For the 15 galaxies in the supplementary sample we have used 21 cm
flux measurements from S05a, as previously mentioned. 
Unlike L07, we have elected to define the ``normal'' galaxy population
for HI content in a manner that is more consistent with our own local
density measurements.\footnotemark 
\footnotetext{We are carrying forward this redefinition of ``normal'' 
HI content to work in progress on the relation between HI deficiency and 
environment.  Hence we wish to establish consistency in this paper with the
future study.}  
The DEF values for these 15 galaxies are based on
$\log(M_{HI})$-$\log(D_{0})$ fits for a large galaxy sample drawn from the
whole Springob et al. (2005a) catalog, which the criterion that
local galaxy densities be less than -1.0 log Mpc$^{-3}$ 
(as described in section 2.1).
We have also chosen a velocity range of 3,000 km s$^{-1}$ to 6,500 
km s$^{-1}$ for selecting a low-density galaxy sample. The lower 
limit is applied to prevent peculiar velocities
from dominating the observed redshift and affecting the line-of-sight 
distance measurements (and, more specifically, to avoid the Virgo
supercluster). We apply an upper velocity limit of 6,500 
km s$^{-1}$ due to the rapidly increasing uncertainty in the LF 
correction to the observed local density (discussed in the previous 
Section), which occurs as the observable LF becomes restricted to 
only the bright end at higher redshift. Of the original $>$8,000 
galaxies in the S05a sample, about 3,000 
galaxies meet the density and 
velocity criteria, which provides a large enough sample to span the necessary
morphological types for fitting the $\log(M_{HI})$- $\log(D_{0})$ relation.
We then perform the same steps described above for calculating the DEF
parameter on the S05a galaxies, where the expected HI mass is now
given by the fits performed on our low density sample.

\begin{figure}
\plotone{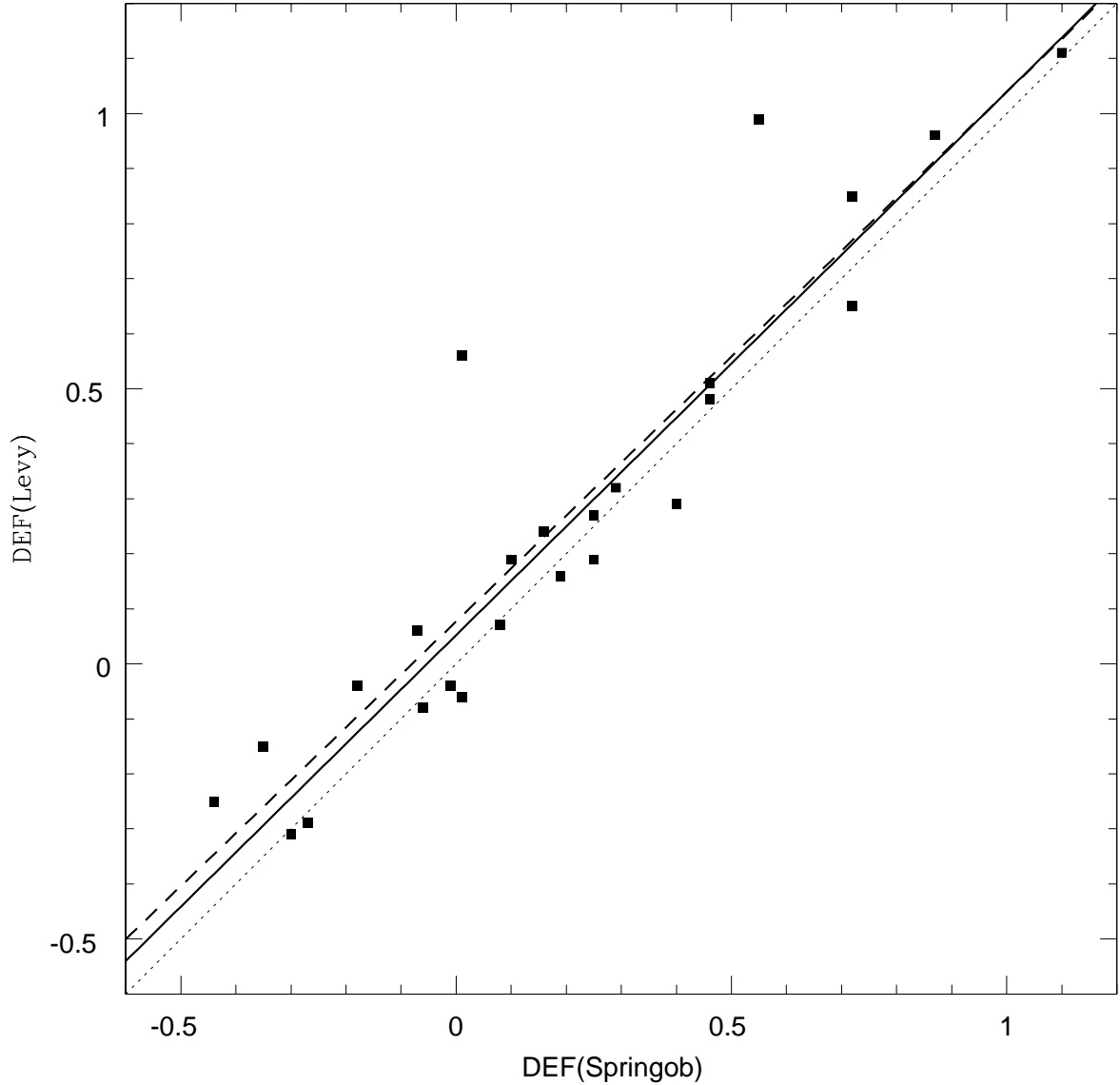}
\caption{The HI deficiency DEF parameter given in L07
(vertical axis) is compared with that determined for this paper from
HI flux data in S05a and our revised prescription for ``normal'' HI
content
(horizontal axis).  The thin dotted line denotes the unit slope line, while
the thick dashed line shows the fit to all 25 galaxies in common between the
two samples, and the thick solid line shows the fit when the most discrepant
point is excluded.}
\label{fig:rose1}
\end{figure}

It is important to verify that 
the HI flux and DEF data in L07 is 
consistently on the same scale as the HI flux data for the 15 galaxies in
the supplementary sample extracted from HI flux data in S05a, and
for which we produced a different definition of ``normal'' HI
content.  To do so, we have extracted
HI flux data for all galaxies in common between L07 and S05a.
For those 25 galaxies we have compared the DEF values given in L07,
which are based on their own Arecibo HI data and on the \citet{so96} 
prescription for ``normal'' HI content, with the DEF values calculated from
the S05a HI data and with the ``normal'' HI content determined from
our new fits to the $\log(M_{HI})$- $\log(D_{0})$ relation.
As seen in Fig.~\ref{fig:rose1}, the agreement between the two data
sources is good.  The thick dashed line shows the linear least
squares fit between the two DEF values for all 25 galaxies in common between
L07 and S05a, while the thick solid line 
indicates the fit achieved when the single most discrepant galaxy, NGC~41, is
excluded.  The slope of the fit is 0.96$\pm$0.08 and 0.99$\pm$0.06 with
and without including NGC~41; the rms scatter about the linear fit is 
$\pm$0.15 and $\pm$0.12 respectively.  We cannot account for the discrepancy
in DEF between the two sources in the case of NGC~41, since the HI fluxes only 
differ by 16\% and the morphological types assumed are the same.  We attribute
the difference to a calculation error in L07.  In the case
of the next most discrepant galaxy, UGC164, the larger DEF value given in
L07 is attributable to the 50\% lower HI flux reported by them.
To maintain consistency with the previous work, we continue to use the
L07 HI flux and DEF data for the galaxies in that paper, with the
exception of NGC~41, for which we adopt the flux data from S05a and
DEF calculation based on the ``normal'' HI content derived for this paper.
Data on DEF for the galaxies in our combined sample
is given in column (2) of Table~\ref{tab:two}.

\begin{figure}
\plotone{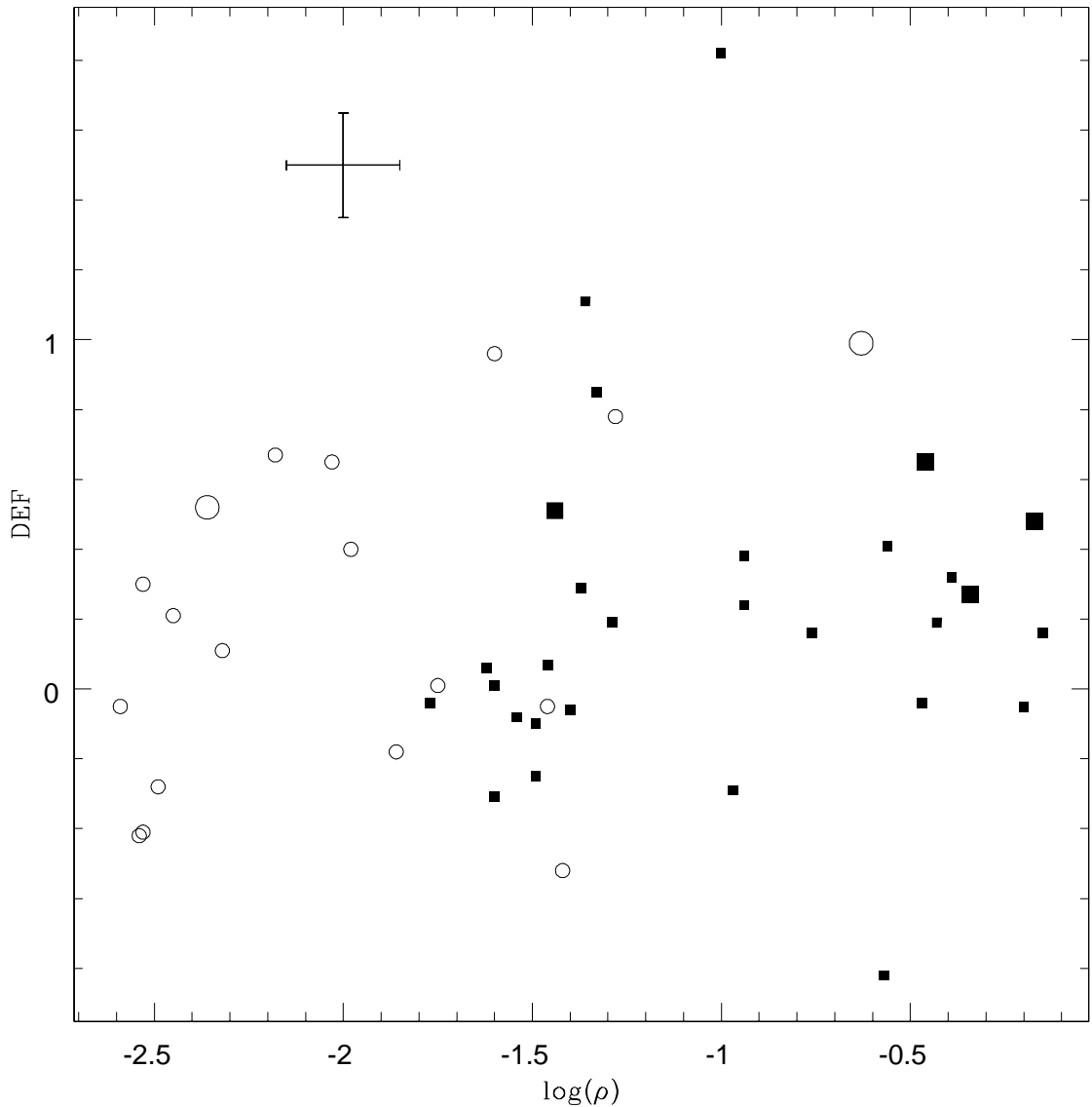}
\caption{The HI DEF parameter (defined in \S2.2) is plotted versus log($\rho$), the local 
environmental density in log Mpc$^{-3}$  
for the data in our combined sample. Filled squares denote Pegasus I cluster 
spirals and open circles are for the
non-cluster spirals.  The larger filled squares and open circles denote
cluster and non-cluster galaxies, respectively, which have both high DEF and 
high central SSFR (i.e., high $\gamma(0.1)$), as discussed in \S3.2.  
Note that the non-cluster spirals were 
preferentially selected to cover a range in DEF at different environments,
hence this figure demonstrates the range of environment and DEF in the sample,
but not a relationship between the two.}
\label{fig:rose2}
\end{figure}

To illustrate the range of $DEF$ and log($\rho$) covered by our Pegasus I and
non-cluster samples, we have plotted $DEF$ versus log($\rho$) in 
Fig.~\ref{fig:rose2}.  The Pegasus I galaxies, plotted as filled squares, on
the whole cover a higher local density regime than the non-cluster galaxies,
plotted as open circles.

\subsection{Optical Imaging}

Broadband B and R and narrowband H$\alpha$ (both online and offline) images
were obtained for 29 galaxies in the Pegasus I cluster and for 18 additional
spiral galaxies with Arecibo HI data with the 4.1-m SOAR telescope atop Cerro
Pachon in Chile during the time period August 2006 through August 2008.  All
images were acquired with the SOAR Optical Imager (SOI).  SOI is a focal
reducing camera mounted at a bent Cassegrain port on SOAR, with a scale of
0.154\arcsc/pixel when read out in 2x2 binned mode \citep{sc04}.  
The 5.3\arcm field of 
view is covered by two E2V 2Kx4K CCDs, with a 7.8\arcs gap between them.  The 
gain is 2.1 e$^-$/ADU and the read noise in binned mode is $\sim$4.5e$^-$.  In
almost every case the galaxy was small enough to fit on a single chip, thereby
avoiding the added complexity of stitching together multiple dithered exposures
to remove the effect of the interchip gap.

The broadband B and R filters are the standard SOI filters, while the H$\alpha$
online and offline filters were fabricated by Custom Scientific, Inc. in 
Phoenix, AZ.  The online filter is centered at 6660 \AA, corresponding to a
redshift of $\sim$3700 \AA \ in the f/9 converging beam, while the offline
filter is centered at 6840 \AA.  Both filters have a FWHM of 100 \AA.

For both the broadband B and R filters a series of three 150 second exposures
was acquired.  For the H$\alpha$ online and offline filters, three 300 second
exposures were taken for each galaxy.  On a typical night 4-6 standard
star fields from \citet{la92} were observed, with typically 3-4 stars 
observed per field.

\subsubsection{Image Analysis}

All images were analyzed with the NOAO/IRAF `mscred' package.  Each raw image
consisted of four image extensions, given that there are two amplifiers on
each CCD chip, and two chips in the SOI instrument.  First, the 
overscan on each amplifier was fit with a linear polynomial and removed, and the data frame was
slightly trimmed.  Next, a master bias frame, combined from a median of 
typically 25 frames, was subtracted from all object and calibration images.
Flat fielding was accomplished by creating a master flat from typically six sky
flat images, taken during evening and/or morning twilight, and with exposure
times adjusted to reach a count level at approximately half of the
linear range of the detector.  The flats were median-combined in the 
`flatcombine' task, with rejection of minimum and maximum values.  Dome flats
were taken as well in each filter every afternoon.  In practice we found that
dome flats and sky flats produced virtually identical results in the R bandpass,
and in the H$\alpha$ online and offline filters, i.e., at red wavelengths.
For the broadband B filter, there were significant differences between dome
and sky flats, due to the plunging spectral energy distribution of the dome
flats in the blue.  Hence sky flats in B were seen to make a far better match to
the object frames, in terms of dividing out obvious structure in the raw images
in scales of tens of pixels.  We consequently used sky flats for all of our
flat fielding.  Dividing through by the sky flat can be problematic if
there is a significant field distortion over the full field of view of the
image.  However, in the case of SOI the maximum scale variation from center
to edge of field is only $\sim$0.15\% at red wavelengths.

At this point the image consisted of four extensions that had been individually
treated.  The next step was to combine the four amplifiers into a single
mosaic image, using an IRAF script written by Dr. Alexandre Oliveira.  Finally,
to combine the three individual exposures in each filter into a single frame
required first registering the images, and then combining them with cosmic
ray rejection.  The tasks `msccmatch' and `wregister' were used in the `mscred'
package to register the images.  The `combine' task, with median scaling and
averaging with the `crreject' cosmic ray rejection option, were used to 
combine the three frames into the final image.  

Since the image quality can differ between the final R band and H$\alpha$ 
online and offline images, the full width half maximum was determined for each
bandpass from several stars in the field.  Then all images were gaussian
convolved to the same FWHM as the worst seeing bandpass.

\subsubsection{Sky Removal and Profile Fitting}

Accuracy of background subtraction typically is the limiting factor in
pursuing galaxy surface photometry at low surface brightness levels.  We
found that even after bias subtraction and flat-field division there is
often a residual discontinuity in the background level across amplifiers on
a chip, and across the boundary between chips.  The discontinuities could be
as large as one ADU per pixel in the broadband R images, out of a background of
approximately 200 ADU, and were usually not constant throughout the night.  
Thus to keep our background subtraction accurate to the 10\% level, we
limited the surface photometry in the R bandpass to a minimum count level
of 10 counts/pixel.  Based on the standard star data for the photometric 
nights, a level of 10 cts/pix corresponds to a surface brightness of
$\sim$24.75 mag~arcsec$^{-2}$ in R at the typical airmass of our
observations.  This surface brightness zero point is consistent with `Lick r'
band
surface photometry given in \citet{co96} for 5 galaxies in common with our
sample.  For those 5 galaxies the radius that we find at the R=24.75 
mag~arcsec$^{-2}$
isophote is on average 7\% smaller than the r=25 mag~arcsec$^{-2}$ isophotal radii
determined by \citet{co96}.  We similarly restricted the H$\alpha$ 
photometry to count
levels brighter than 0.5 cts/pix in the online and offline filters.

Sky background was found from the median count levels in regions on the
same chip as the galaxy that were found to be free of bright objects.  These
regions were defined interactively on the image and the mean value for the
multiple regions was used as the final background level.

Once the sky background was removed from the H$\alpha$ online and offline 
images, aperture photometry was performed on several stars.  The offline
image was normalized to attain the same mean stellar counts as in the 
online image.  The final H$\alpha$ image is the difference between the
online and normalized offline images.

For almost all of the finalized R band images, we used the `ellipse' task in
the `stsdas/isophote' IRAF package to determine an azimuthally averaged
radial R band profile for each galaxy.  The `ellipse' task starts with initial
approximations to the x-y center, position angle, and ellipticity of the 
isophote at a specified radial distance, then, iterates to optimum values, then
steps along to both larger and smaller radii.  Among the results tabulated at
each azimuthally averaged radius are the surface brightness level, 
in counts/pix and in magnitudes, the total magnitude within the radius, and 
errors on all quantities.  As mentioned above, we cut off our analysis at a
surface brightness level of 10 cts/pix, corresponding to a surface
brightness in R of $\sim$24.75 mag/sq. arcsec.  In a few cases the galaxy
was either so close to edge-on, or so irregular in morphology, that profile
fitting could not be accomplished reliably.  For those galaxies we instead used
the `polyphot' task to interactively draw a polygonal boundary around the
galaxy, within which the total magnitude was integrated.  Thus in these cases
only a total R magnitude was determined, as opposed to radial profiles.  With
a little experimentation using both `polyphot' and `ellipse' tasks, we found
that by eye a polygon could readily be drawn around a galaxy which returned the
same global magnitude as produced from `ellipse' at the 10 cts/pix isophote.

The H$\alpha$ image is naturally less subject to stable isophote fitting than
the R band image.  Consequently, for each galaxy we applied the x-y center,
PA, and ellipticity information determined from the R band profile fitting 
to the H$\alpha$ image.  In the case of those galaxies for which only global
R band magnitudes were attained with `polyphot', the same polygon was applied to
the H$\alpha$ image.  However, there were a few galaxies with just a few low
luminosity HII regions.  For those cases we carried out aperture photometry
of the individual HII regions, and then calculated a combined H$\alpha$ flux. 
The H$\alpha$ flux includes as well the emission in the neighboring
[NII]$\lambda$6548,6584 forbidden lines.

\subsubsection{Measurement of Specific Star Formation Rates}

From the above analysis we extracted the following information.  First, we
define the parameter $\Gamma$,
a global characterization of the star formation rate per unit stellar mass, 
by taking the base 10 logarithm of the total H$\alpha$ flux within the
0.5 cts/pix isophote divided by the total R band flux within the 10 cts/pix 
isophote, as follows:

\begin{equation}\Gamma=log\left(\frac{f_{H\alpha}}{f_R}\right)\end{equation}

The $\Gamma$ parameter is essentially equivalent to the ``normalized massive
star formation rate'' discussed in KK04ab.  
The H$\alpha$ flux characterizes the current star formation rate
(SFR) of massive stars, while the R band flux is a rough characterization of the total stellar
mass, assuming that the mass to R band light ratio is roughly constant in our
disk galaxy sample.  Since both the H$\alpha$ and R band data were acquired within a 
short enough time period that the airmass of the galaxy did not change
substantially, and since the R band and H$\alpha$ filters are at similar
wavelengths, the raw flux ratio did not require correction for extinction
effects.   We henceforth refer to this normalized SFR as the {\it specific}
star formation rate, or SSFR.  Note that our version of the SSFR is slightly
different than that used by other authors because we are not basing it on
flux-calibrated H$\alpha$ and R band photometry.  Thus the zeropoint of our
SSFR values will differ from that of other investigators, but the relative
values should be on the same scale.

As pointed out by FG08, there is a systematic luminosity
dependence in the SFR per H band luminosity, i.e., more luminous galaxies have
a lower SSFR.  FG08 found
a slope to that relation of -0.26 in the log of the SFR per unit H band
luminosity versus the log of the H band luminosity, with a substantial
($\sim$0.3 mag) scatter.  To account for this
trend in our analysis, we have applied the FG08 correction as
follows.  We calculated the $M_H$ for all galaxies in our sample using the
total H band magnitude and distance modulus given by Hyperleda.  In cases
where Hyperleda provides only the total B magnitude, we assumed an average
$B-H$ color of 2.52, determined from the mean of our sample for which data on
both passbands is available.  We then defined a corrected version of the
log of the H$\alpha$ to R band flux, $\Gamma^*$, given by:

\begin{equation}\Gamma^*=\Gamma+0.26[-0.4(M_H+22.50)],\end{equation}

\noindent where the correction is defined to be zero for an absolute H magnitude
of -22.50, which is the mean for our sample.  Data on $\Gamma$, $\Gamma^*$,
and $M_H$ are given in columns (3), (4), and (5) of Table ~\ref{tab:two}.
The radius, in arcseconds, containing the azimuthally averaged surface
brightness level 10 counts/pixel, $r_{10}$, is given in column (6) of
Table ~\ref{tab:two}.  The ratio of the H$\alpha$ radius
(i.e., to the 0.5 cts/pix level) to $r_{10}$, $r_{H\alpha}/r_{10}$, is listed in
column (7). The half-light effective radius, $r_{eff}$, which is determined
from the surface brightness profile, is given in 
column (8).  

Errors in the $\Gamma$ and $\Gamma^*$ values are difficult to assess, given that
the photon statistical errors are exceedingly small, thus systematic errors
dominate.  We have identified two primary sources of systematic uncertainty,
namely, (1) uncertainties in establishing the mean sky level and (2)
uncertainties in fitting non-axisymmetric disk structure with ellipses and
low order deviations.  We evaluated the first uncertainty by experimenting with
different sky levels, and the second with comparing isophotal fit fluxes with
those extracted from subjectively defined enclosed polygons (using the IRAF
'polyphot') task.  We estimate the $\pm$1$\sigma$ errors in $\Gamma$ and 
$\Gamma^*$ (and in the later defined $\Gamma_{H\alpha}$, $\Delta$, and
$\gamma(0.1)$ quantities) to be $\sim\pm$0.15 mag.  The errors are larger for
the lower surface brightness galaxies with faint and irregular HII regions.
We estimate the 
$\pm$1$\sigma$ errors in $r_{H\alpha}/r_{10}$ to be typically $\sim\pm$0.2.
The error in DEF is estimated to be $\sim\pm$0.15.

\begin{figure}
\plotone{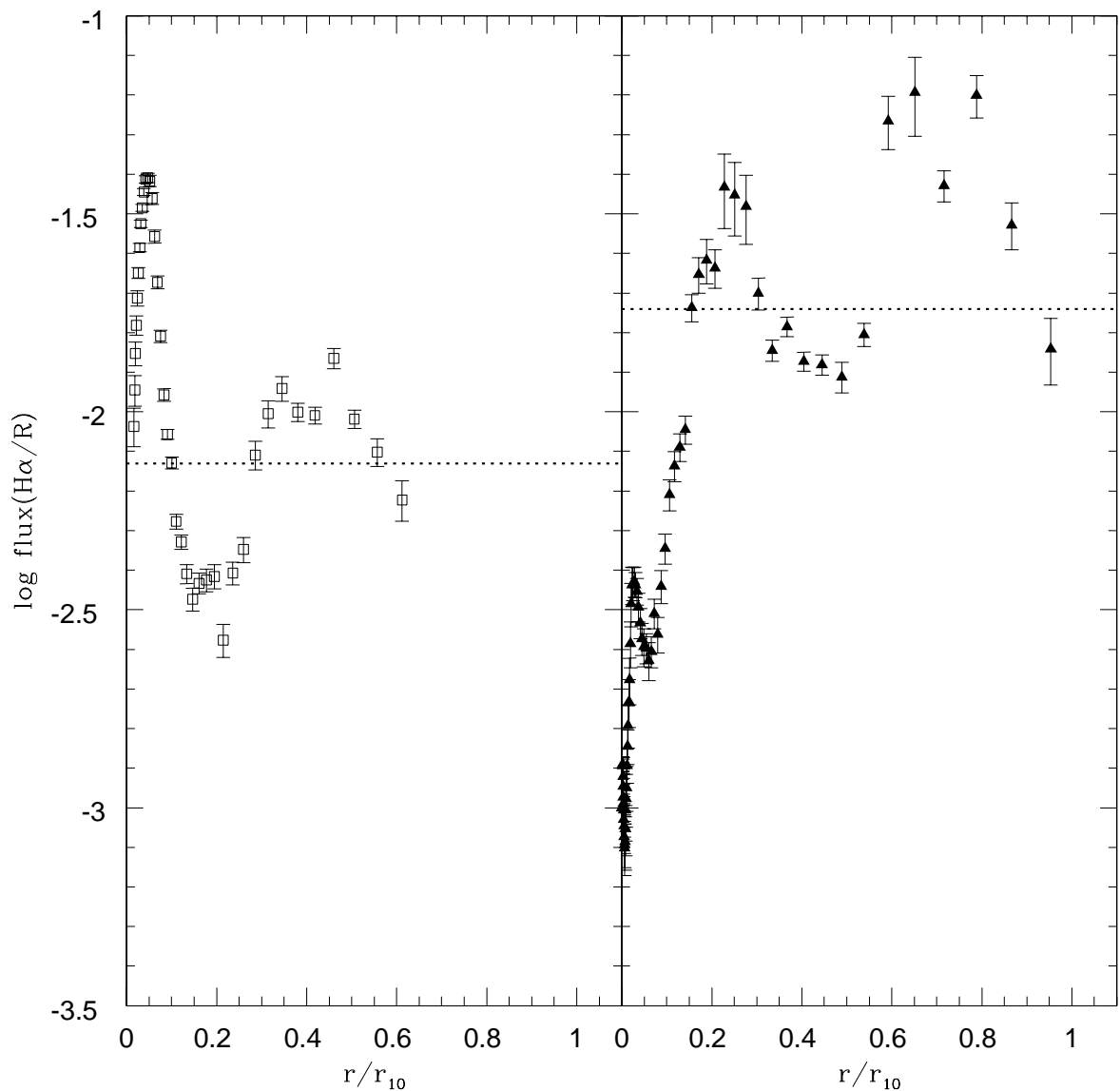}
\caption{Azimuthally averaged radial profiles for the logarithm
of the flux ratio in H$\alpha$ to that in the R bandpass are plotted against
normalized radial distance.  The radial distance is normalized by the radius of
the 10 cts/pix isophote.  The galaxy in the left panel, NGC7518, is HI 
deficient, with DEF=0.27, while the galaxy in the right panel, UGC11524, is
HI-rich, with DEF=-0.52.  The error bars correspond to the 1$\sigma$ errors in 
both the R and H$\alpha$ fluxes, combined in quadrature.  The horizontal dashed
line in each panel represents the global value of the H$\alpha$ to R band flux
ratio for that galaxy.}
\label{fig:rose3}
\end{figure}

\begin{figure}
\plotone{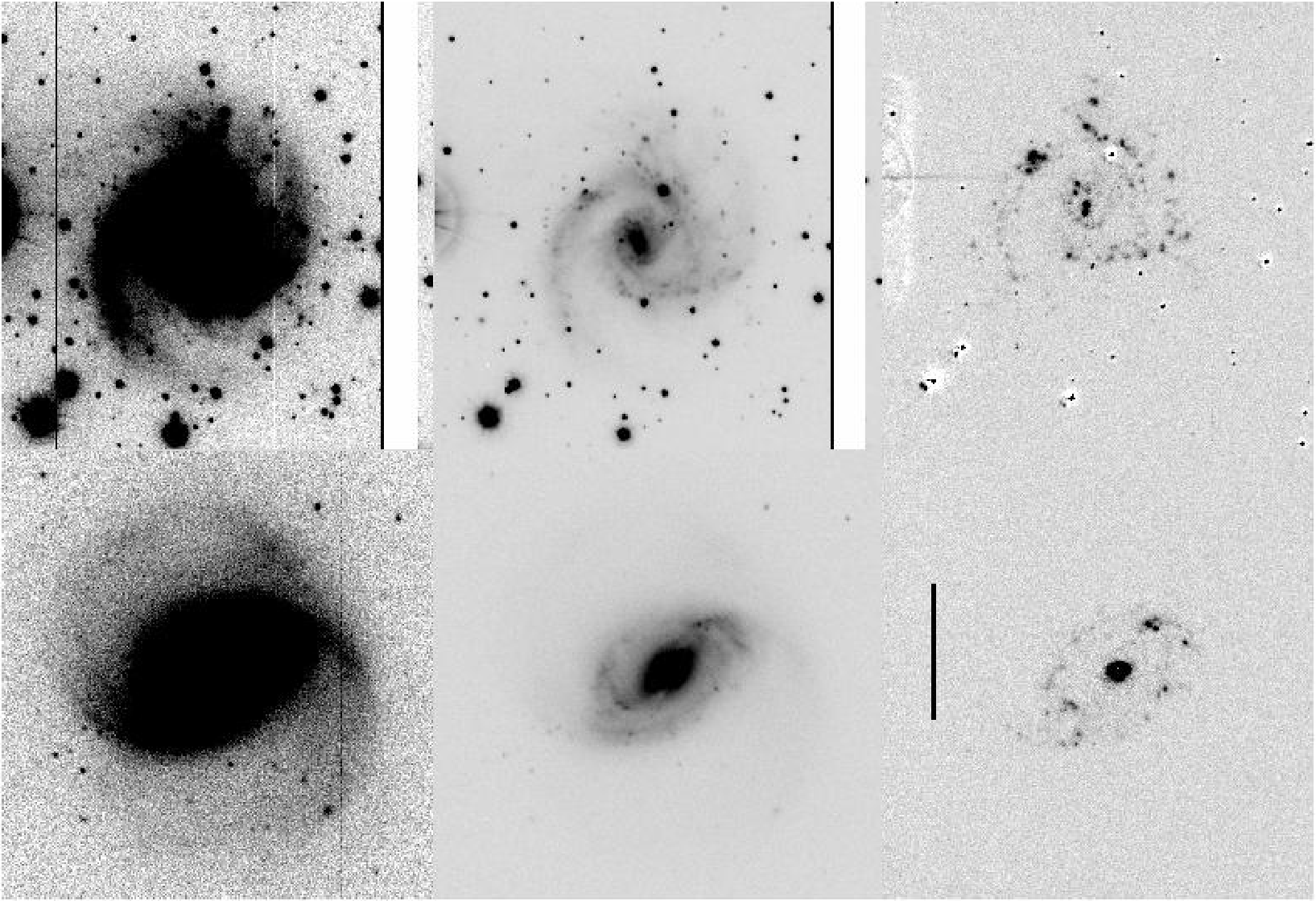}
\caption{R band and H$\alpha$ images are displayed for the gas-rich spiral
UGC11524 (upper panels) and for the gas-depleted spiral NGC~7518 (lower
panels).  The R band images are shown at two contrast levels in the left
and middle panels, while the H$\alpha$ images are shown in the right panels.
The black line in the lower right panel indicates an angular scale of 30\arcsc.}
\label{fig:rose4}
\end{figure}

In addition to the global SSFR, we produced the radial profile of
the log of the H$\alpha$ to R band flux ratio from the azimuthally averaged
ellipse fitting.  We refer to this radially dependent quantity as $\gamma(r)$.
In Fig.~\ref{fig:rose3} radial profiles are illustrated
for both an HI-depleted galaxy (NGC7518) in the left panel and an HI-rich
galaxy (UGC11524) in the right panel.  The differences in the radial profiles 
of the two galaxies is discussed later.  In Fig.~\ref{fig:rose4} R band and H$\alpha$ images
are shown for both galaxies.

\subsection{Optical Spectroscopy}

Optical spectra of three HI-deficient galaxies were obtained with the Goodman
Spectrograph at the SOAR telescope on the night of September 14-25, 2008.  These
three galaxies are examples of HI-deficient spirals with high central SSFR that
are defined and discussed in \S3.2.  The
Goodman Spectrograph is an all-refracting optics imaging spectrograph mounted 
at one of the Nasmyth foci, and utilizes volume phase holographic gratings as 
the dispersing element \citep{cl04}.  The spectra are imaged onto a 4k x 4k
Fairchild 486 back-illuminated CCD.  For our observations the spectra were binned by 2 pixels in the dispersion direction and by 4 pixels along the slit.
With the 600 lines/mm VPH grating the binned pixels sample the spectrum at 
1.3 \AA/pix, at a resolution of 7 \AA \ FWHM, and cover the spectral range
4350-7000 \AA.  A 1.65\arcs x 3.9\arcm slit was used at the North-South
position angle for all observations.  Wavelength calibration was achieved with
a HgCuAr lamp.  For each galaxy a pair of 300 second exposures was acquired.

\section{Results}

The primary goal of this paper is to investigate whether HI deficiency
in spiral galaxies is also reflected in star formation rates.  Specifically,
we first investigate whether a global HI deficiency in a spiral disk produces
a corresponding global suppression in SSFR.  We then examine our radial
SSFR profiles to assess whether galaxies with global HI deficiencies
exhibit characteristic signatures in the radial distribution in SSFR,
for instance, truncation of the star forming disk.

\subsection{Global SFR}

\begin{figure}
\plotone{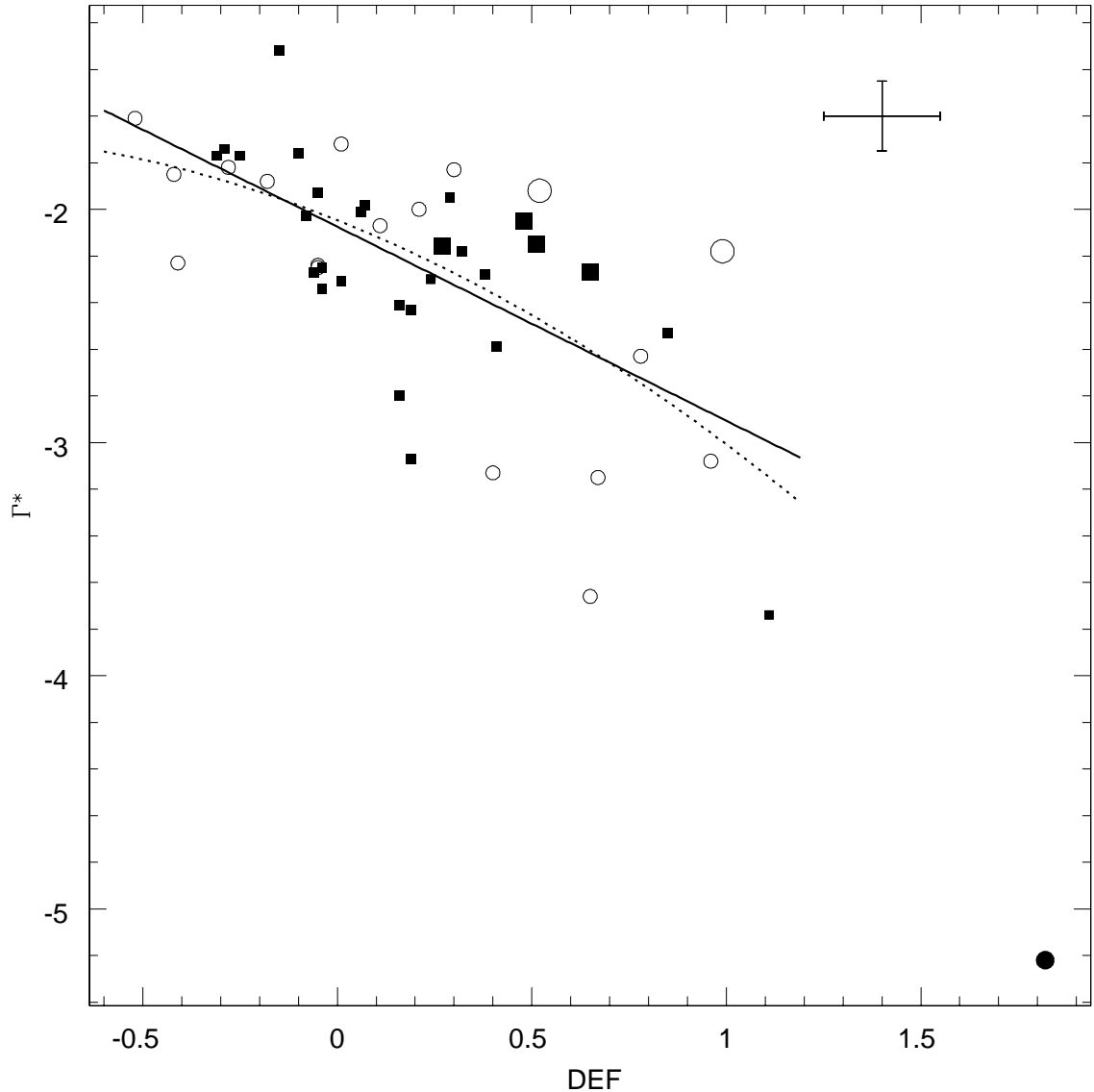}
\caption{The corrected global SSFR parameter, $\Gamma^*$
(defined in \S2.3.3), 
is plotted versus the gas deficiency parameter, DEF, for the data in our
combined sample. Filled squares denote Pegasus I cluster spirals and open 
circles are for the
non-cluster spirals.  The larger filled squares and open circles denote
cluster and non-cluster galaxies, respectively, which have both high DEF and 
high central SSFR (i.e., high $\gamma(0.1)$), as discussed in \S3.2.  The linear
least squares fit is shown as a solid line, while the quadratic fit is shown as
the dotted curve.  The large filled circle represents an upper limit in $\Gamma^*$
and a lower limit in DEF for the galaxy NGC~7563, and is not included in the 
fits.  The error bar in the upper right corner shows typical $\pm$1$\sigma$
errors.}
\label{fig:rose5}
\end{figure}

To assess whether a global deficit in atomic hydrogen in spirals is connected 
with the global SSFR, in Fig.~\ref{fig:rose5} we have plotted the SSFR
parameter, $\Gamma^*$,
versus the HI DEF paramter.  The large filled circle denotes the
lower limit in DEF and upper limit in $\Gamma^*$ for NGC~7563. 
As previously discussed, $\Gamma^*$
is a measure of the SSFR, corrected for 
the correlation between SSFR and total H band luminosity.  As seen
in Fig.~\ref{fig:rose5}, there is a trend toward lower SSFR with
increasing HI deficit.  A linear least squares fit between $\Gamma^*$ and DEF,
which does not include NGC~7563,
yields a correlation coefficient r=-0.68, with rms scatter of 0.46; the 
probability that the two variables are uncorrelated is rejected at the 
2 x 10$^{-7}$ level.  Thus $\Gamma^*$ and DEF are clearly related.  The linear 
fit is denoted by the solid line in
Fig.~\ref{fig:rose5}, and is plotted only over the DEF range for which it
is constrained.  We have also plotted the quadratic fit (dotted line),
which yields an rms scatter of 0.45.  The slope of the linear fit is
-1.05$\pm$0.17.  Replacing $\Gamma^*$ with the ``uncorrected'' 
$\Gamma=log\frac{f_{H\alpha}}{f_R}$ produces virtually identical results.
As previously mentioned, FG08 found a substantial intrinsic scatter in their
relationship between SSFR and L$_H$, thus a considerable intrinsic scatter
is present in the global SSFR of galaxies, in addition to any dependance on
$DEF$ or environment.

While the quadratic fit indicates a flattening of the relation between
$\Gamma^*$ and DEF at lower DEF, i.e., for gas-rich galaxies, it does not 
significantly lower the rms residual. Hence, it would appear as if 
a linear fit is sufficient to decribe the relation, with the
implication that the global SSFR tracks the global HI even into the gas-rich 
regime.  However, a closer look at Fig.~\ref{fig:rose5} indicates that the 
scatter around both linear and quadratic fits is dominated by the large 
intrinsic scatter at high DEF, and thus cannot be improved by increasing the 
order of the fit.  We consequently divided the sample into high (DEF$\ge$0.0)
and low (DEF$<$0.0) galaxies.  We made separate linear least squares fits to
the two samples, which yields a slope of -1.22$\pm$0.31 and rms residual of
$\pm$0.52 for the 30 high DEF galaxies, and a correlation coefficient 
of r=-0.60.  The hypothesis that there is no correlation between $\Gamma^*$ 
and DEF in the high DEF sample is rejected at the 0.00048 level.
For the 17 galaxies in the low DEF sample, we obtain a slope of 
-0.79$\pm$0.57 and rms residual of $\pm$0.33.  With a correlation coefficient of
r=-0.34 the hypothesis of no correlation is only rejected at the 0.18 level.
However, if the one discrepant galaxy with $\Gamma^*$$>$-1 is excluded, we
then obtain a slope of -0.92$\pm$0.32 and rms scatter of only $\pm$0.18.
The correlation coefficient is r=-0.61, and the no correlation hypothesis is
rejected at the 0.011 level.

To summarize, there is a clear correlation between $\Gamma^*$ and DEF, with
SSFR decreasing with increasing HI deficiency.
This result confirms that found for Virgo spirals in KK04ab and FG08.
If the one very high $\Gamma^*$ galaxy at low DEF (i.e., gas-rich) is excluded,
then the trend between $\Gamma^*$ and DEF is significant at greater than the
3$\sigma$ level at both high (DEF$\geq$0.0) and low (DEF$<$0.0) DEF.  
While the slope derived for low
DEF galaxies is formally lower than for high DEF galaxies, the difference is
not statistically significant.  Hence our main conclusion is that the relation
between $\Gamma^*$ and DEF extends from gas-rich to gas-poor spirals.  We
also find that the rms scatter around the relation is substantially higher for 
the high DEF (gas-poor) galaxies.

\subsection{SFR Profiles}

We now consider whether in high DEF galaxies the SSFR is suppressed throughout
the galaxy, or whether it is preferentially lower (or higher) at particular 
locations in the disk.  As is discussed in \S4, this question has been 
addressed by KK04ab and FG08 for Virgo cluster spirals, with some disagreement
between the two studies.
Given that conclusions regarding the relative importance of starvation versus
stripping phenomena can be arrived at from the characteristic depletion
patterns in both SSFR and HI, the detailed radial profiles in $\gamma$(r) are
worth a careful examination.

To begin with we return to Fig.~\ref{fig:rose3}, in which the radial profiles
in $\gamma$(r) are plotted versus the azimuthally averaged isophotal radius,
normalized by the radius at the 10 cts/pix isophote, R$_{10}$, for both a 
gas-poor (NGC~7518) and a gas-rich (UGC11524) galaxy.  As can be seen in 
Fig.~\ref{fig:rose3}, the
$\gamma$(r) profiles are complex, and not readily described by a simple radial
gradient in the SSFR.  What is most striking about
the two profiles is that the HI depleted galaxy actually has a high SSFR in
its central region, while it is low everywhere else.  On the other hand, 
$\gamma$(r) is actually low in the central region of the HI-rich galaxy, but
is high everywhere else.  The contrast between H$\alpha$ and R band light in
NGC~7518 versus in UGC11524 is further seen in the H$\alpha$ and R band images
shown in Fig.~\ref{fig:rose4}.

To quantify the contrast in central SSFR between the high and low DEF galaxies,
we have defined the quantity $\Delta$, as follows:

\begin{equation}\Delta = \gamma(0.1) - \Gamma,\end{equation}

\noindent where $\gamma(0.1)$ is the mean of the 10 highest values of
$\gamma$(r) found in the central $0.1 R_{10}$ radius of the galaxy.  Thus
$\Delta$ represents the difference between this highest central value of
$\gamma$(r) and the global value $\Gamma$.  A positive value of $\Delta$
indicates that the SSFR is higher in the center than it is
globally; the global value is the average out to the 10 cts/pix radius
R$_{10}$.  

\begin{figure}
\plotone{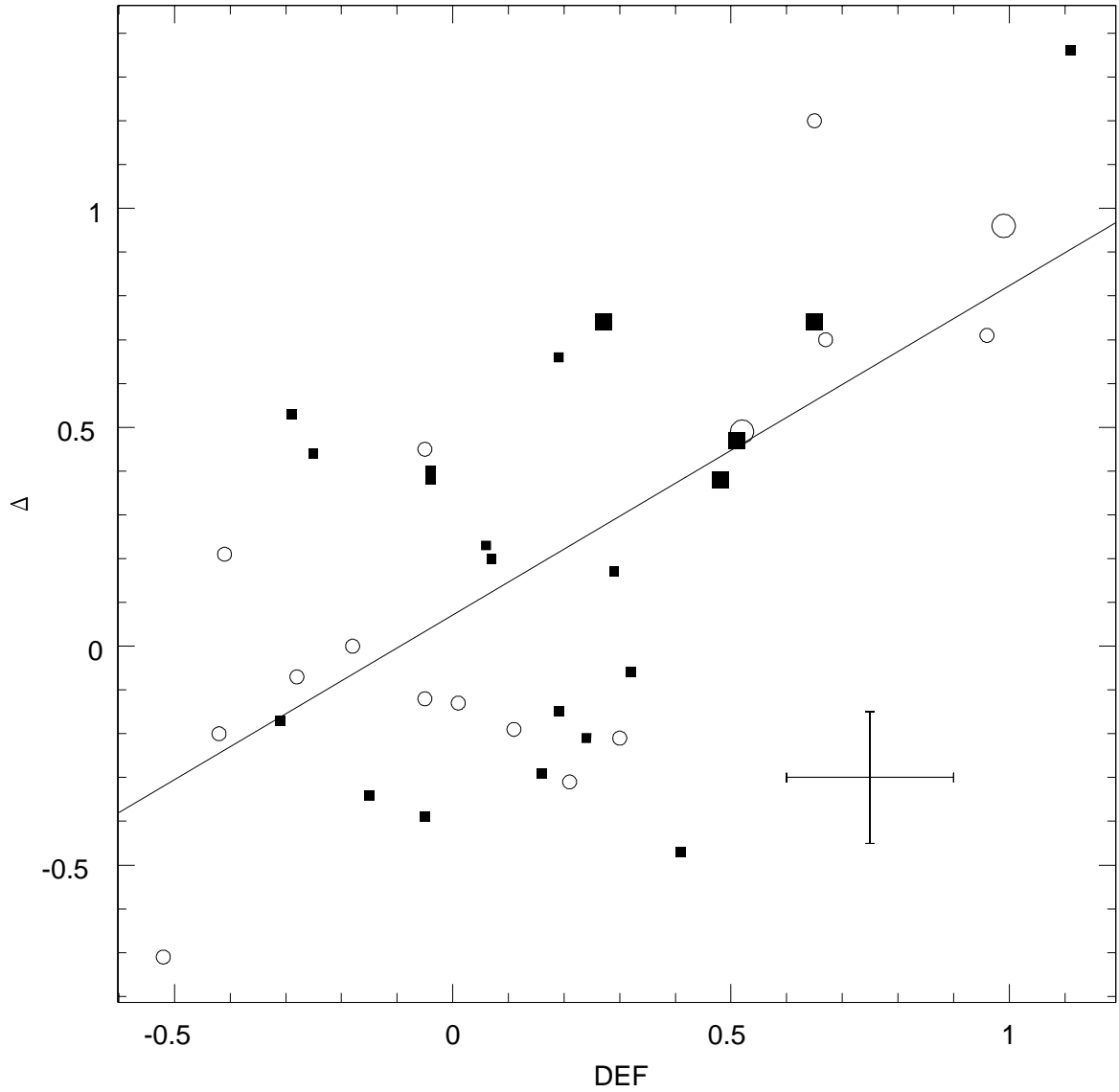}
\caption{The $\Delta$ parameter, representing the difference in logarithmic
SSFR between the central and global values
(defined in \S3.2), is plotted
versus DEF for the galaxies in our sample with measurable values for $\Delta$.
The least squares fit to the data is denoted by the solid line.  All symbols
are the same as in Fig.~\ref{fig:rose5}.}
\label{fig:rose6}
\end{figure}

We have determined $\Delta$ for the 37 galaxies in which 10 valid data points are
available within $0.1 R_{10}$; the $\Delta$ values are given in column (10)
of Table~\ref{tab:two}, while the $\gamma(0.1)$ values are in column (9).
In Fig. \ref{fig:rose6} the $\Delta$ values are plotted versus DEF.  
In addition to
a considerable scatter in the plot, a significant correlation 
between $\Delta$ and DEF is found.  Specifically, a linear least squares fit 
yields a slope of 0.79$\pm$0.17
and the correlation coefficient is 0.66.  The hypothesis that the two
variables are uncorrelated is rejected at the 3.5 x 10$^{-5}$ level.  

A perhaps
surprising result is that the scatter around the fit between $\Delta$ and
DEF is actually lower at high DEF than it is at low DEF.  This is in contrast
to our previous examination of the fit between global $\Gamma^*$ and DEF,
where the scatter around the fit at high DEF is greater than at low DEF.  In
Fig. \ref{fig:rose6} there appears to be a transition at DEF$\sim$0.3 such
that at DEF$\gtrsim$0.3 all galaxies have a high value of $\Delta$, i.e.,
centrally concentrated star formation, and the scatter around a linear trend
between $\Delta$ and DEF is small.  For DEF$<$0.3, no significant trend
is found between $\Delta$ and DEF, and the scatter in $\Delta$ is nearly a
factor of 2 greater.

\begin{figure}
\plotone{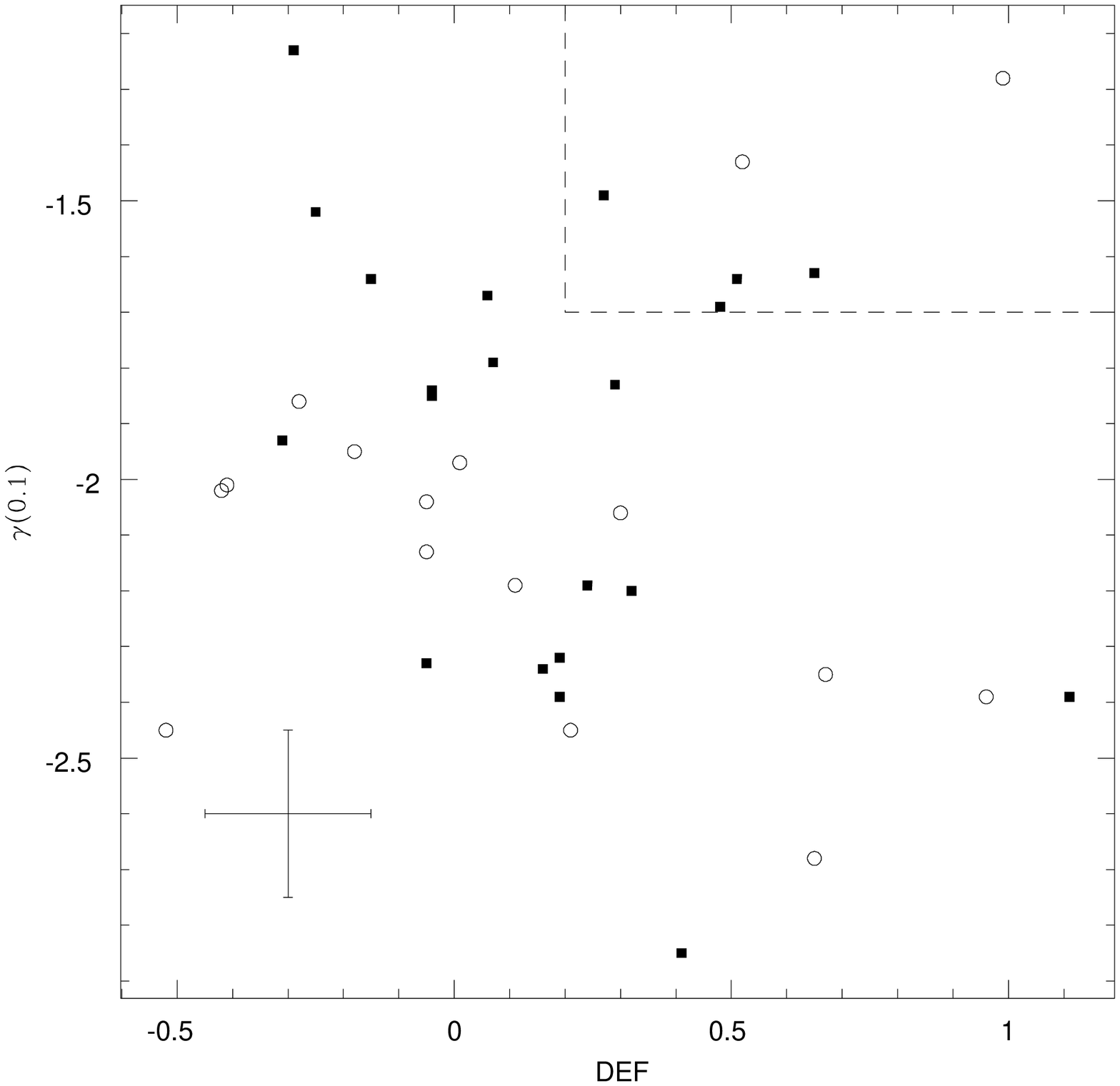}
\caption{The maximum central SSFR parameter, $\gamma(0.1)$
(defined in \S3.2),
is plotted versus DEF for the galaxies in our sample.  Filled squares and open circles are
for Pegasus I cluster and non-cluster galaxies, respectively.  
The dashed box region in
the upper right corner isolates the HI deficient galaxies with high peak
central SSFR.}
\label{fig:rose7}
\end{figure}

An additional look at the central SSFRs in our sample is provided in 
Fig.~\ref{fig:rose7}, where $\gamma(0.1)$ is plotted against DEF, i.e., 
we simply plot the $\gamma(0.1)$ measure of the highest
central SSFR, without differencing it with the global value.  While 
in Fig.~\ref{fig:rose6} {\it all} galaxies with DEF$\gtrsim$0.3 have high 
$\Delta$ values, in Fig.~\ref{fig:rose7}
the high DEF galaxies appear to bifurcate into a high $\gamma$(0.1) and a low
$\gamma$(0.1) group.  The high $\gamma$(0.1) group of 6 galaxies, that lie 
within the dashed region in the upper right of Fig.~\ref{fig:rose7}, 
have higher $\gamma(0.1)$ values than nearly all of the low DEF
galaxies, while the opposite is true for the low $\gamma(0.1)$ group. 
Remarkably, then, the HI-deficient high $\gamma(0.1)$ galaxies actually have 
higher central SSFR than typical gas-rich galaxies.  While the 
remaining low $\gamma(0.1)$ galaxies at high DEF have lower central SSFR 
than gas-rich spirals, their SSFR is so low globally that they
still have high $\Delta$ values.

\begin{figure}
\plotone{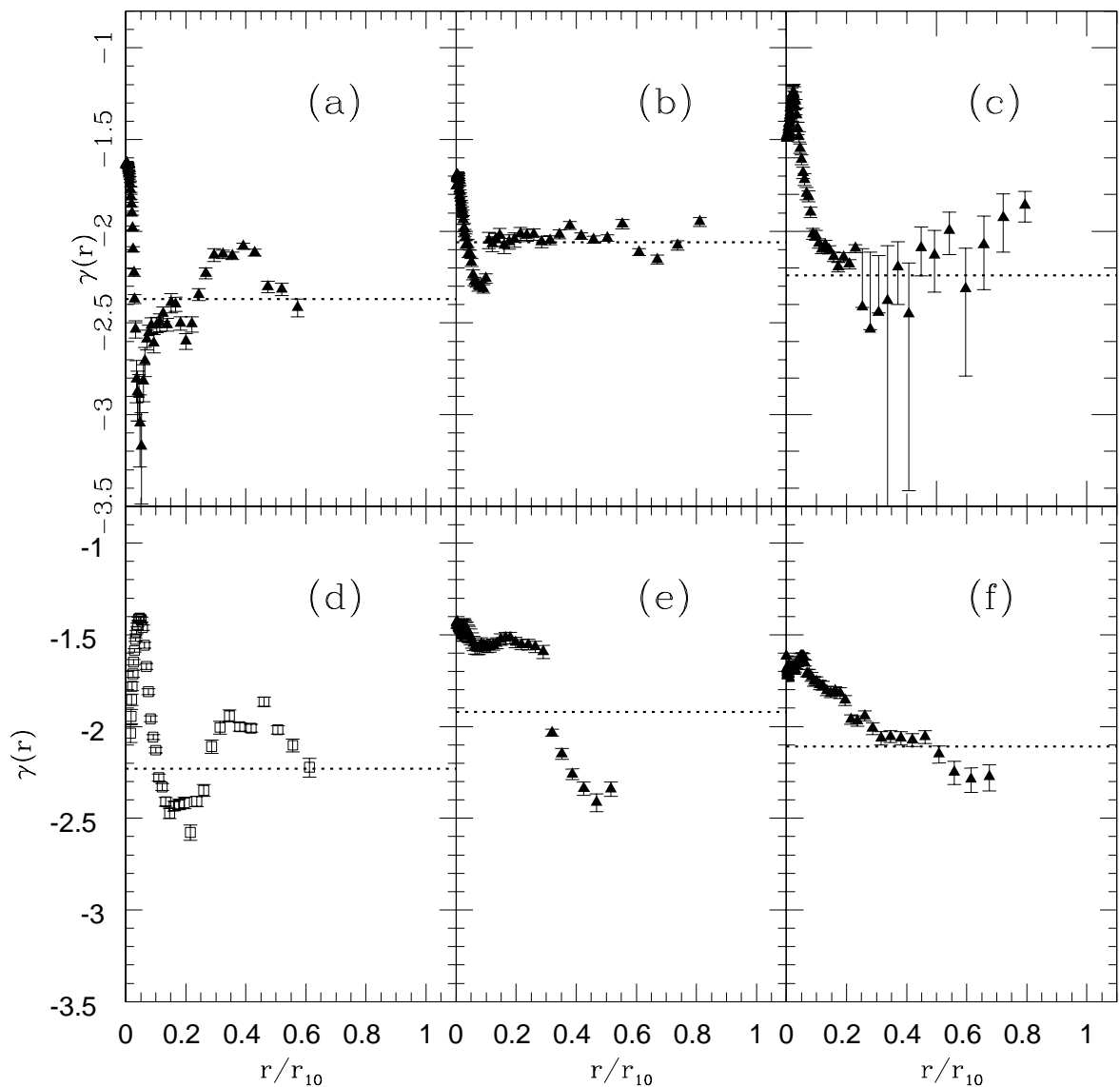}
\caption{Radial $\gamma$(r) SSFR profiles are plotted for the six high DEF
galaxies with high central SSFR, $\gamma$(0.1).  The six galaxies are:
(a) NGC~7518, (b) NGC~7608, (c) UGC164, (d) NGC~7643, (e) NGC~7074, and
(f) UGC12535. The error bars correspond to the 1$\sigma$ errors in
both the R and H$\alpha$ fluxes, combined in quadrature.  The horizontal dashed
line in each panel represents the global value of the H$\alpha$ to R band flux
ratio for that galaxy.
}
\label{fig:rose8}
\end{figure}

\begin{figure}
\plotone{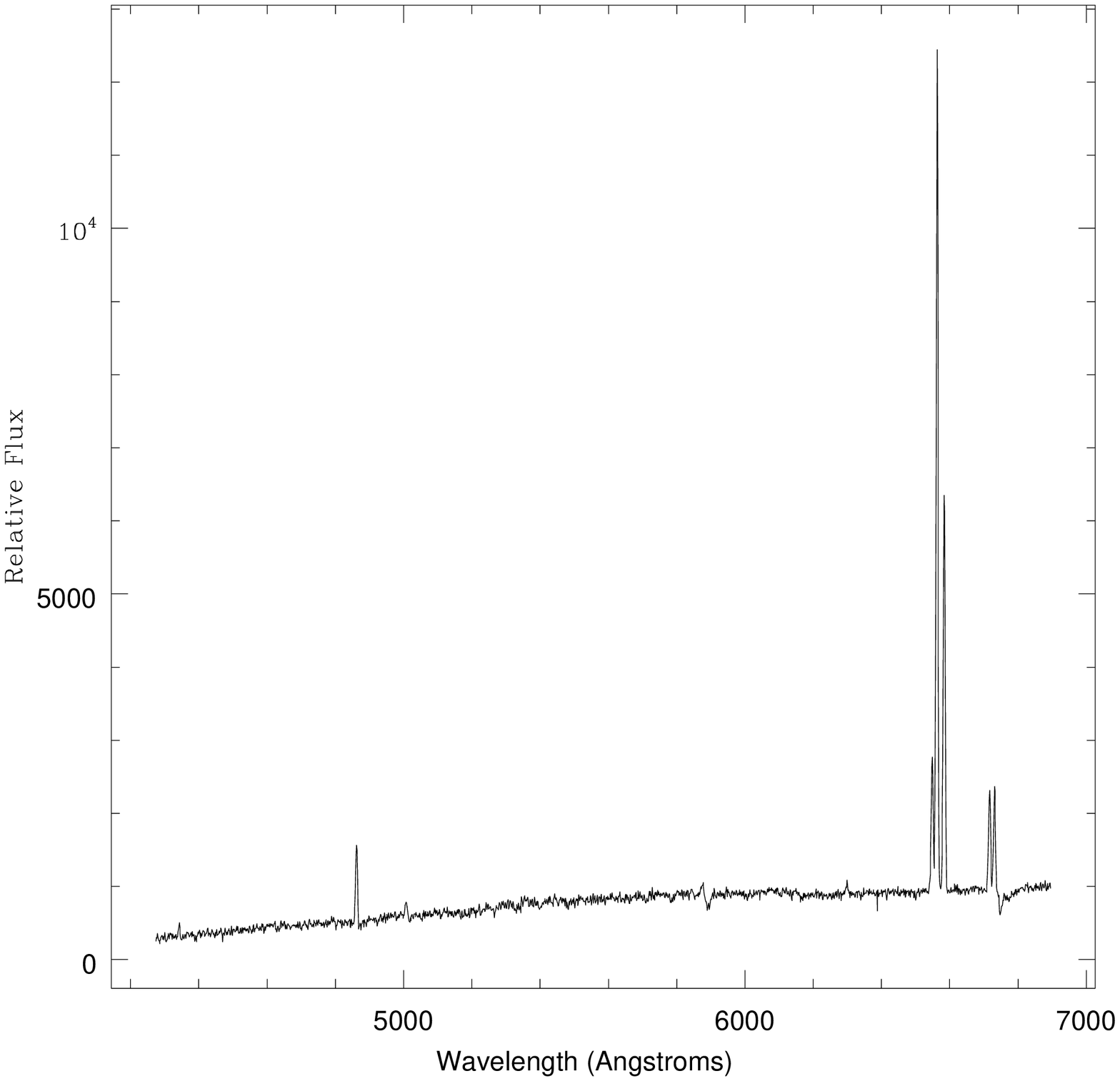}
\caption{Deredshifted optical spectrum of UGC164.}
\label{fig:rose9}
\end{figure}

The radial $\gamma$(r) SSFR profiles for the 6 high $\gamma(0.1)$ gas-poor
galaxies are shown in Fig.~\ref{fig:rose8}.  With the exceptions of
NGC~7074 and UGC12535, the central region of elevated SSFR is highly
centrally concentrated.  However, without emission line diagnostics, one
cannot assume that the H$\alpha$ (and [NII]$\lambda$6548,6584) emission is
due to star formation, as opposed to an active galactic nucleus (AGN).
As is discussed in \S2.4 we acquired optical spectra of three of the
high $\gamma(0.1)$ galaxies, NGC~7518, NGC~7643, and UGC164, covering
both H$\alpha$ and H$\beta$, including the [NII]$\lambda$6548,8584 and 
[OIII]$\lambda$4959,5007 lines near  H$\alpha$ and H$\beta$ respectively.
The line ratios [OIII]$\lambda$5007/H$\beta$ and [NII]$\lambda$6584/H$\alpha$
are useful diagnostics for distinguishing between AGN/LINER spectra and HII
region spectra due to star formation (e.g., Baldwin, Phillips, \&
Terlevich 1981).  In Fig.~\ref{fig:rose9} the nuclear spectrum of
UGC164 is plotted.  While H$\alpha$ is more than a factor of 2 greater than
[NII]$\lambda$6584, the excitation is very low, i.e., [OIII]$\lambda$5007 is
barely detected, while H$\beta$ in emission clearly rises well above the 
underlying H$\beta$ absorption in the integrated stellar spectrum.  The
combination of low excitation ([OIII]$\lambda$5007/H$\beta$=0.17)
and H$\alpha$ approximately twice [NII]$\lambda$6584 
unambiguously distinguishes the nuclear emission spectrum of UGC164 as
that of a metal-rich HII region in Fig.~5 of Baldwin, Phillips, and
Terlevich (1981).  A similar conclusion is arrived at for NGC~7518
and NGC~7643.  Thus for the three high DEF spirals with centrally concentrated
H$\alpha$ emission for which we have optical spectra, we conclude that the
emission is indeed due to high levels of star formation, rather than from an
AGN.

\begin{figure}
\plotone{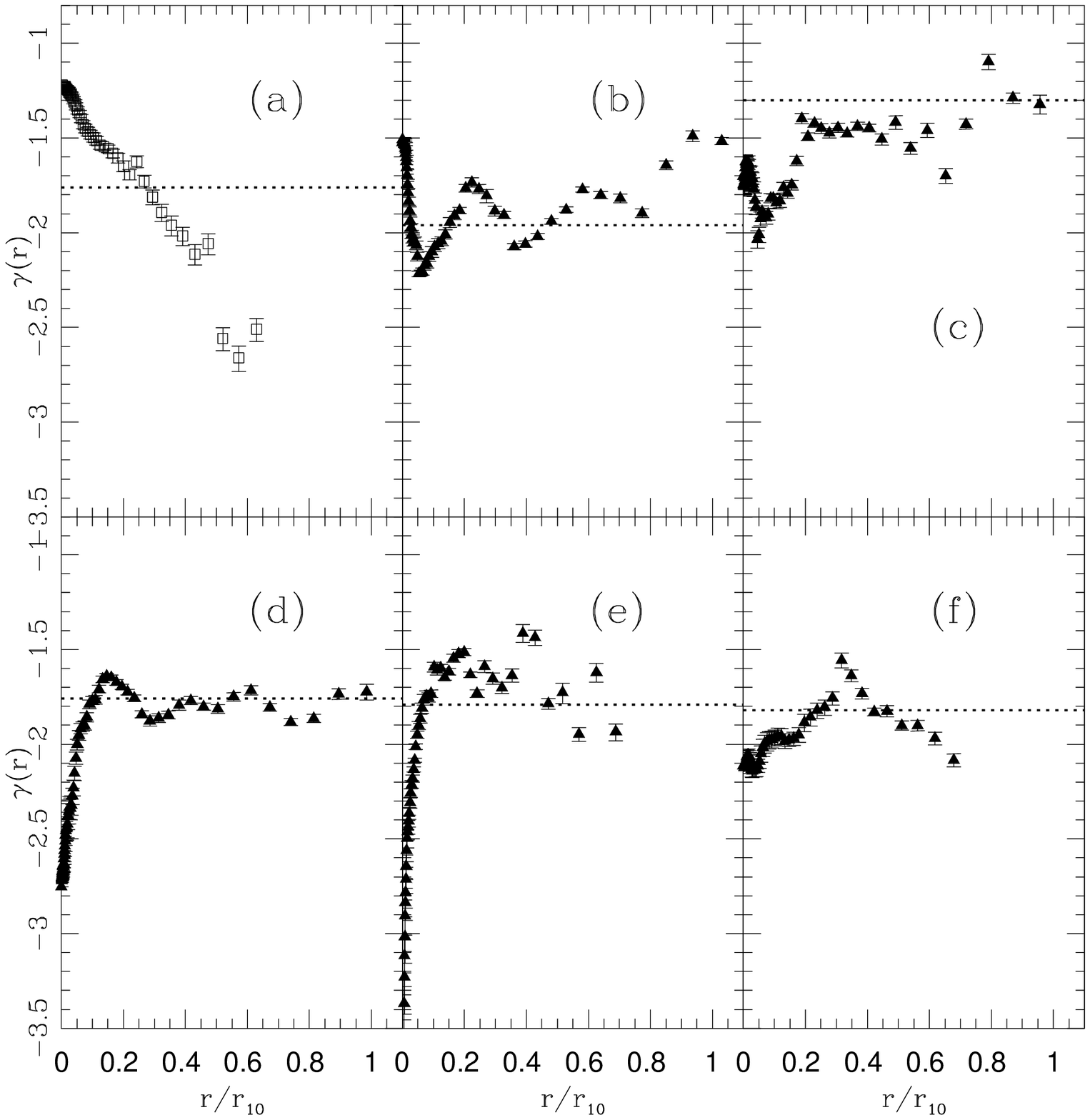}
\caption{Radial $\gamma$(r) SSFR profiles are plotted for three low DEF
(gas-rich) spirals with high central SFR in the upper three panels:
(a) NGC~7580, (b) NGC~7591, and (c) NGC~7610.  Low DEF spirals with
more typical central SFR of $\gamma$(0.1)$\sim$-2 are plotted in the
lower panel: (d) IC~1205, (e) UGC11759, and (f) NGC~7537.
The error bars correspond to the 1$\sigma$ errors in
both the R and H$\alpha$ fluxes, combined in quadrature.  The horizontal dashed
line in each panel represents the global value of the H$\alpha$ to R band flux
ratio for that galaxy.
}
\label{fig:rose10}
\end{figure}

To obtain further perspective on the high DEF galaxies with high $\gamma(0.1)$,
we compare their radial $\gamma(r)$ profiles with those of gas-rich
galaxies having a variety of $\gamma(0.1)$ values.
In Fig.~\ref{fig:rose10} are plotted $\gamma(r)$ profiles for gas-rich
galaxies with both high central SSFR of $\gamma$(0.1)$>$-1.7 (upper panel) 
and ``typical'' central SSFR of $\gamma$(0.1)$\sim$-2 (lower panel).  The
profile for NGC~7580 (Fig.~\ref{fig:rose10}a) shows the more spatially extended high SSFR that is
also seen in two of the six gas-poor high $\gamma(0.1)$ galaxies,
NGC~7074 and UGC12535.  NGC~7580 is a Markarian galaxy, i.e., UV-excess
on low-dispersion objective prism plates (Markarian 1967), and is included
in a list of starburst nucleus galaxies in Coziol (2003).  While neither
NGC~7074 and UGC12535 are Markarian galaxies, NGC~7074 is described as a
``post-eruptive'' triplet in Zwicky (1971).  In contrast, NGC~7591
(Fig.~\ref{fig:rose10}(b)) has a centrally concentrated SSFR peak, is also a
Markarian galaxy, but its optical emission-line spectrum is characterized
by Veron, Goncalves, and Veron-Cetty (1997) as composite LINER and HII region. 
NGC~7610 (Fig.~\ref{fig:rose10}(c)) is seen to have a high SSFR core, but
more striking is the high $\gamma(r)$ throughout the disk.  Its nuclear
emission spectrum has been characterized as AGN by Schombert (1998).

Finally, we note that if the high DEF galaxies with high $\gamma(0.1)$, i.e.,
the galaxies with high absolute central SSFRs, are removed from 
consideration, the correlation between $\Delta$ and DEF is still present,
with a slope of 0.68, a correlation coefficient of 0.55, and a probability
that the two parameters are uncorrelated rejected at the 0.0015 level.  Thus
the correlation between $\Delta$ and DEF plotted in Fig.~\ref{fig:rose6} is
not just an effect produced by the high DEF galaxies with high $\gamma(0,1)$.

In summary, gas-poor galaxies have higher SSFR in their centers relative to
the average disk value (i.e., higher $\Delta$), than do gas-rich galaxies.  
In fact, approximately half of the gas-poor galaxies have higher central 
SSFR (i.e., higher $\gamma(0.1)$) than is typical of the 
gas-rich galaxies.  Optical spectroscopy of several of the gas-poor spirals
with high central SSFR indeed establishes that the H$\alpha$
emssion is due to SF, rather than to an AGN.  

\subsection{H$\alpha$ Disk Truncation}

\begin{figure}
\plotone{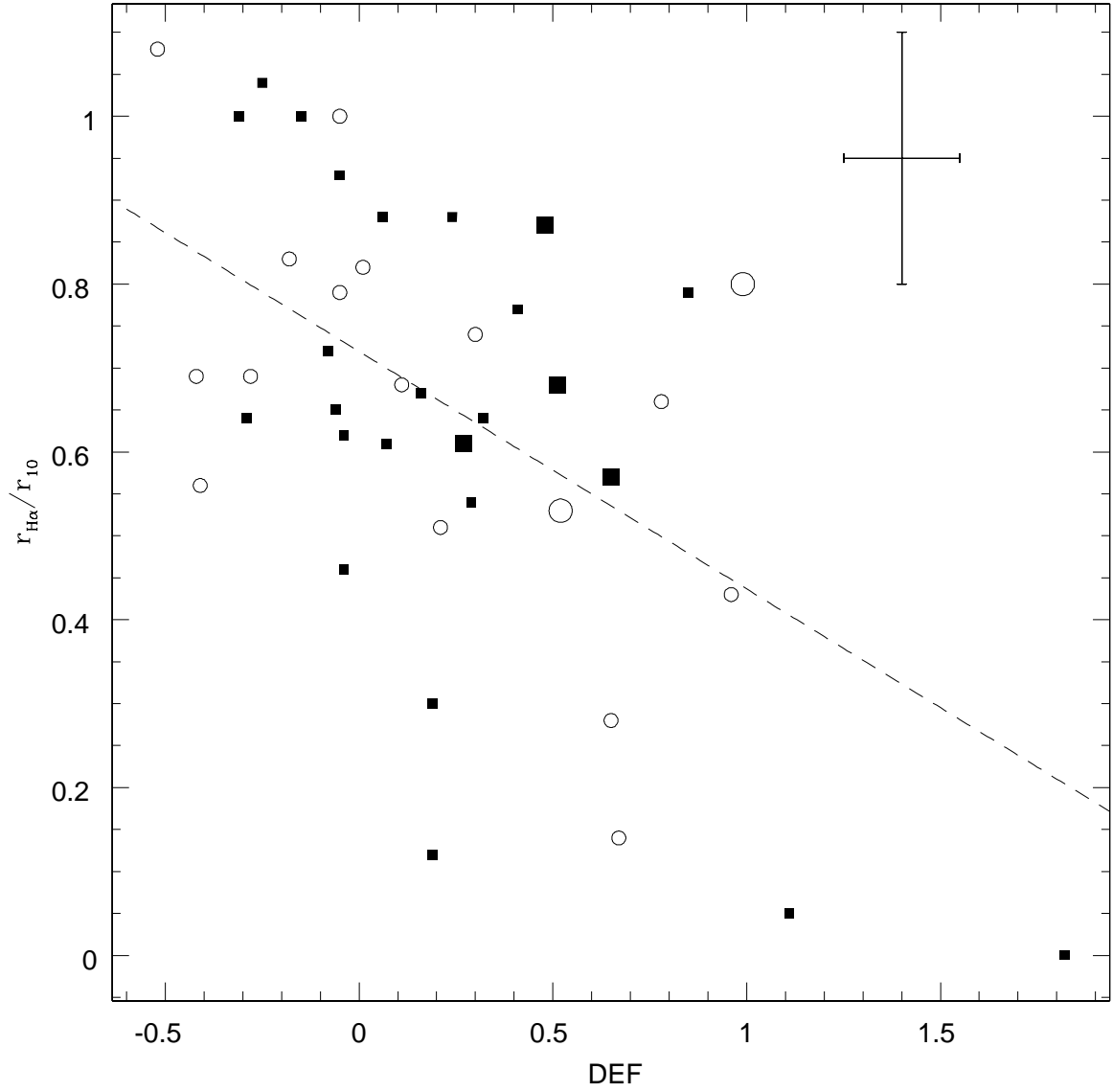}
\caption{The H$\alpha$-to-R band disk size ratio is plotted versus DEF.  The
dashed line represents the linear least squares fit to the data.  All symbols
are the same as in Fig.~\ref{fig:rose5}.}
\label{fig:rose11}
\end{figure}

The preceding Section \S3.2 on radial SSFR profiles has emphasized the elevated
{\it central}
H$\alpha$ levels found in many high DEF galaxies (in an otherwise low SSFR
and gas-depleted disk).  KK04b find a correlation between the size of the 
H$\alpha$ disk and that of the HI disk, with high DEF galaxies exhibiting 
truncated disks. Defining the radial extent of something as knotty as an
H$\alpha$ flux distribution is challenging.  In KK04b the H$\alpha$ disk
length is defined as the radius containing 95\% of the H$\alpha$ flux.  Here
we define the H$\alpha$ disk length to be the outermost radius at which the
H$\alpha$ surface brightness is above the 0.5 cts/pix level (that is
mentioned in \S2.3.2 to be the flux level above which we can reliably trace the
emission).  We divide this H$\alpha$ disk radius by the R band radius, r$_{10}$,
to create a normalized H$\alpha$ disk size, $r_{H\alpha}/r_{10}$.  The
H$\alpha$-to-optical disk sizes are plotted versus DEF in 
Fig.~\ref{fig:rose11}, along with the linear least squares fit.  The fit
has slope of -0.28$\pm$0.08, a correlation coefficient r=-0.48, and rejects
the hypothesis of no correlation at the 0.0016 level.  Hence, we find a 
significant correlation between the H$\alpha$-to-optical disk size,
$r_{H\alpha}/r_{10}$, and HI deficiency, DEF,
indicating a progressive radial truncation of the star-forming disk with
increasing HI deficiency.

\begin{figure}
\plotone{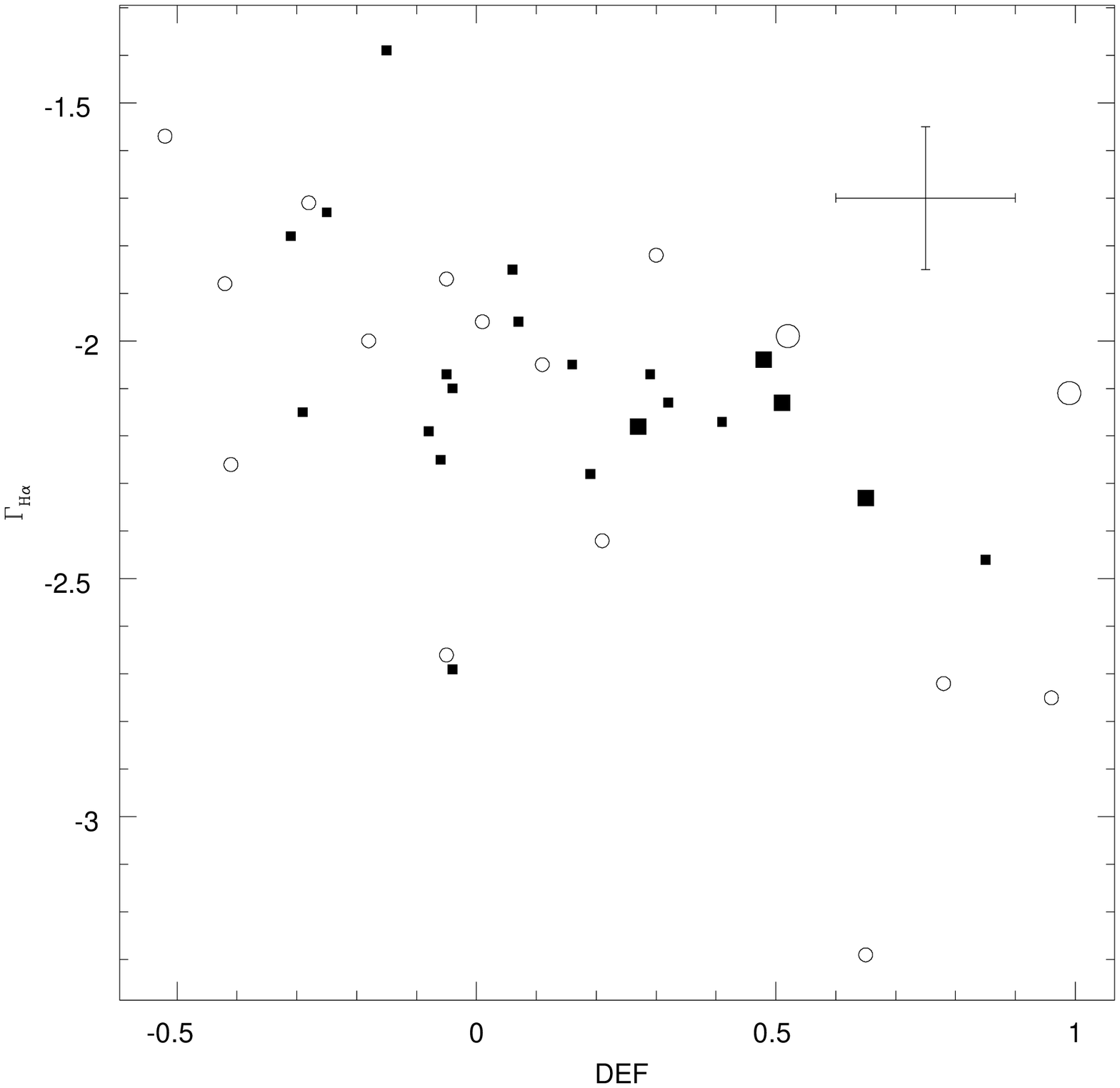}
\caption{The average SSFR within the star-forming part of the disk, 
$\Gamma_{H\alpha}$
(defined in \S3.3), is plotted versus DEF.  All symbols
are the same as in Fig.~\ref{fig:rose5}
}
\label{fig:rose12}
\end{figure}

\begin{figure}
\plotone{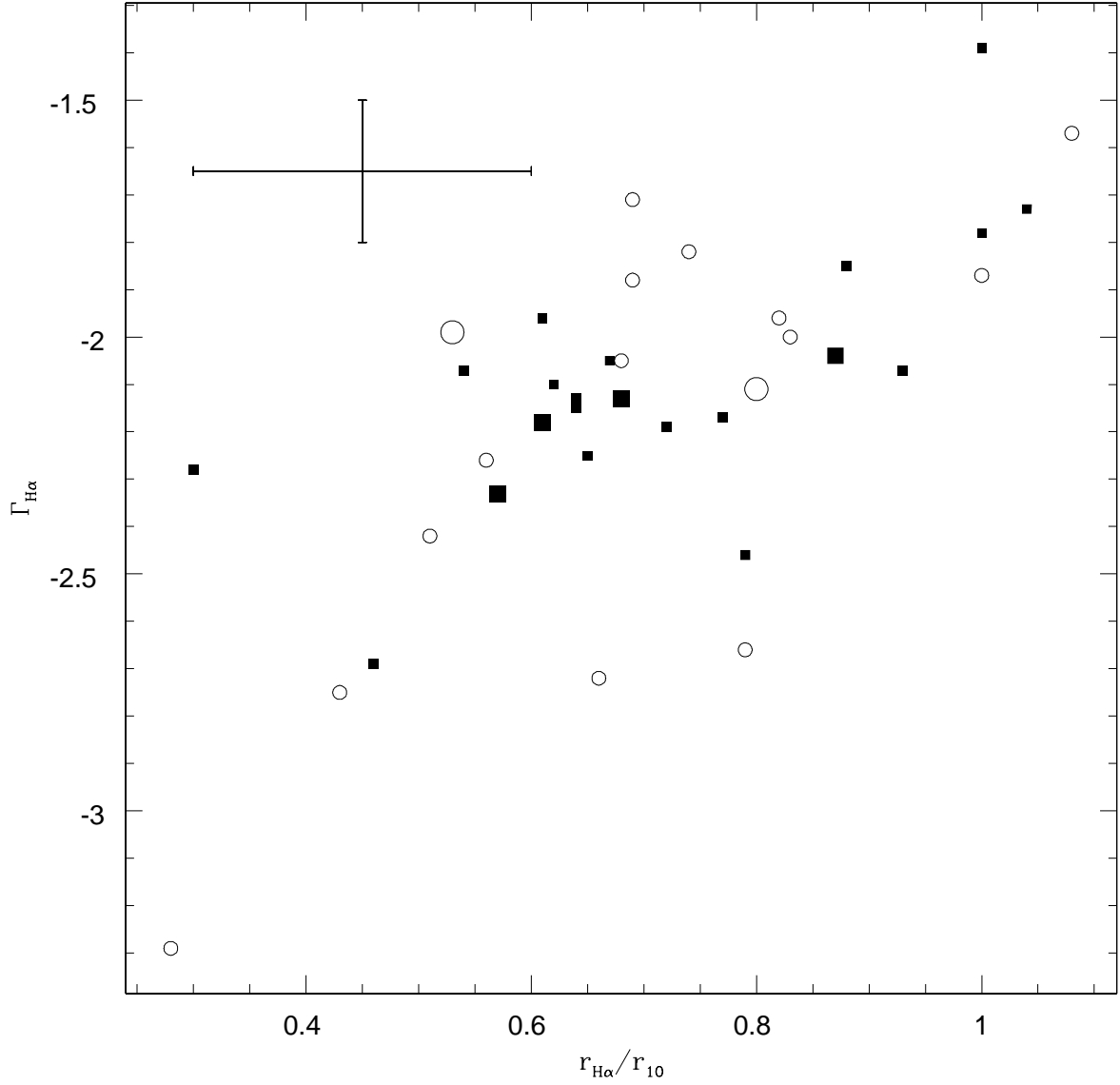}
\caption{The average SSFR within the star-forming part of the disk,
$\Gamma_{H\alpha}$, is plotted versus, $r_{H\alpha}/r_{10}$,
the H$\alpha$-to-R band disk size ratio.  All symbols
are the same as in Fig.~\ref{fig:rose5}.
}
\label{fig:rose13}
\end{figure}

An additional question relating to the truncated H$\alpha$ disks is whether or
not the SSFR in the active part of the disk is at normal levels.  That is, are
the disks producing stars at normal rates out to the radius at which
they are are truncated?  To address this question we calculated the average
$\gamma(r)$ within the radial region $0.2r_{10}<r<r_{H\alpha}$.  Note that we
exclude the inner $0.2r_{10}$ to avoid being affected by the high central SSFR
in many of the gas-poor galaxies, thereby ensuring an unbiased average of the
disk (as opposed to nuclear) SSFR.  This average disk SSFR, $\Gamma_{H\alpha}$
(which is listed in column (11) of Table~\ref{tab:two} ), is plotted in
Fig.~\ref{fig:rose12} and is well correlated 
with DEF; the slope of the relation is -0.49$\pm$0.13 (the hypothesis of no
correlation is rejected at the 0.006 level).  In Fig.~\ref{fig:rose13} 
$\Gamma_{H\alpha}$ is plotted against $r_{H\alpha}/r_{10}$, which reveals that
these two parameters are highly correlated; the hypothesis of no correlation
is rejected at the $2.6 \times 10^{-6}$ level.  
Hence we can affirm that the local
SSFR is suppressed in truncated disks.

A comparison of the slopes of the correlations between $\Gamma^*$ and DEF
(slope of  -1.05$\pm$0.17), between $r_{H\alpha}/r_{10}$ and DEF (slope of
-0.28$\pm$0.08), and between $\Gamma_{H\alpha}$ and DEF (slope of
-0.49$\pm$0.13), further reveals that 
local suppression in and truncation of the star forming disk both 
contribute equally to the global correlation between SSFR and DEF.  Since
SSFR$\sim$$(r_{H\alpha}/r_{10})^2$, then the fact that the square of the
slope of the 
$r_{H\alpha}/r_{10}$ $vs$ DEF correlation, $2(-0.28)=-0.56$, is within
the errors equal to the -0.49 slope of the correlation between $\Gamma_{H\alpha}$
and DEF, indicates that the two trends contribute equally to the correlation
between global SSFR and DEF.  In addition, we can assume that
SSFR is proportional to $\Gamma_{H\alpha} (r_{H\alpha}/r_{10})^2$. Thus the
fact that the observed slope of $\Gamma^*$ $vs$ DEF, -1.06, is within
the errors equal to -0.49+2(-0.28) indicates that we have a single population of
galaxies in which the star forming disks are {\it both} truncated and suppressed, rather
than two populations, one truncated and the other suppressed.

\subsection{Morphology of the High Central SFR Galaxies}

While the preceding examination of the azimuthally averaged radial 
SSFR profiles provides quantitative data concerning the nature of HI deficient
spirals, further insight on these galaxies is in principle contained in their
detailed morphologies.  We first return to Fig.~\ref{fig:rose4}, in which the R band and
H$\alpha$ images of NGC~7518, a prototype for the HI-deficient galaxies with
high central SSFR, are contrasted with those of UGC11524, a prototype for the
gas-rich spirals.  The H$\alpha$ image of NGC~7518 is clearly characterized by 
a bright emission core, and a ring of HII regions which are coincident with the
end of the bar seen in the middle panel R band image.  Furthermore, the 
H$\alpha$ emission is truncated at the end of the bar, while 
the faint outer R band light is seen to extend well beyond
the H$\alpha$ disk.  In contrast, for UGC11524 there is no bright H$\alpha$
emission core.  In addition, the H$\alpha$ emission traces the R band light
even to the faint outer spiral arms.

\begin{figure}
\plotone{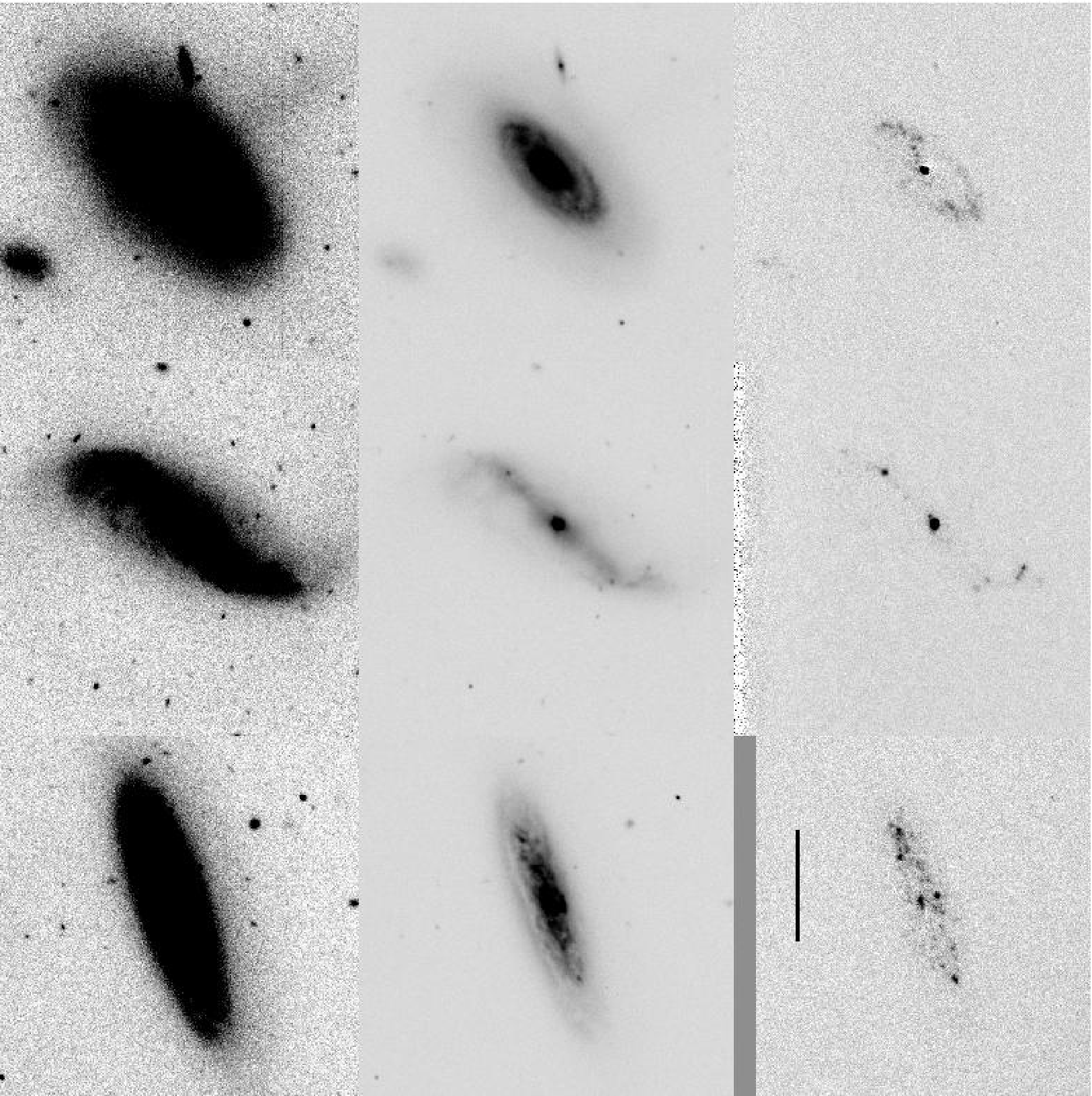}
\caption{R band and H$\alpha$ images are displayed for three gas-poor spirals
with bright central H$\alpha$ emission cores.  NGC~7643 is in the upper
panels, UGC164 in the middle panels, and NGC~7608 is in the lower panels.
The R band images are shown at two contrast levels in the left
and middle panels, while the H$\alpha$ image are shown in the right panels.
The black line in the lower right panel indicates an angular scale of 30\arcsc.}
\label{fig:rose14}
\end{figure}

Two of the three other high DEF galaxies with a central H$\alpha$ emission
core exhibit similar morphologies to NGC~7518, as is seen in Fig.~\ref{fig:rose14}.
NGC~7643 in particular has a similar morphology, i.e., an emission core,
a ring of HII regions that coincide with the end of a bar structure, and then
a faint R band disk that extends well beyond the H$\alpha$ disk.  UGC164 is
similar as well, with a very bright H$\alpha$ emission core, and a clear stellar
bar structure, with no H$\alpha$ emission evident beyond the bar.  The faint
R band light beyond the bar is difficult to characterize, however.  The fourth
galaxy, NGC~7608, is too close to edge-on to tell whether a bar structure is
present, and whether the extranuclear HII regions form a ring.  

\begin{figure}
\plotone{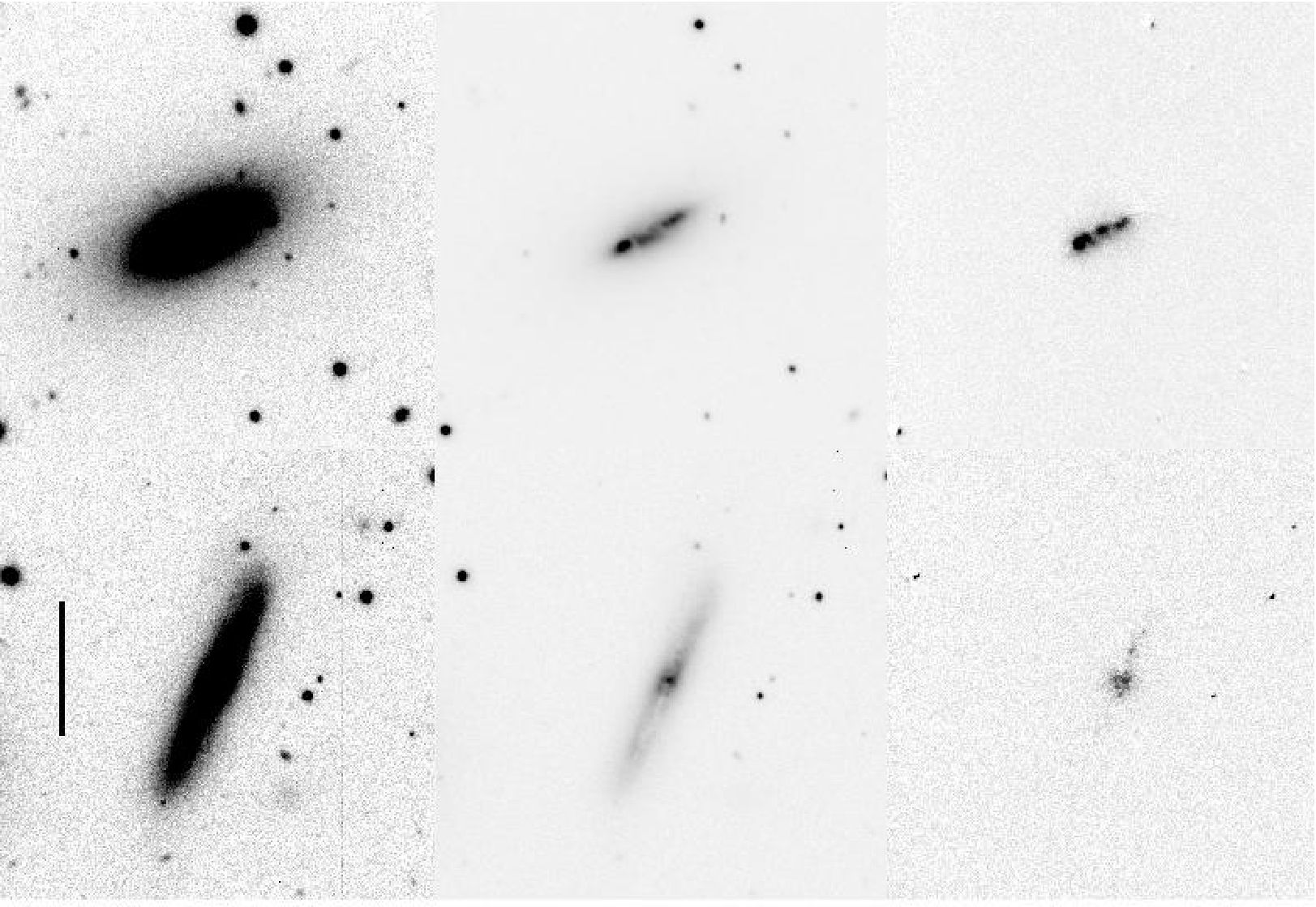}
\caption{R band and H$\alpha$ images are displayed for two gas-poor spirals
with extended bright central H$\alpha$ emission regions.  NGC~7074 is in the
upper panel and UGC12535 is in the lower panel.
The R band images are shown at two contrast levels in the left
and middle panels, while the H$\alpha$ images are shown in the right panels.
The black line in the lower left panel indicates an angular scale of 30\arcsc.}
\label{fig:rose15}
\end{figure}

Two other high DEF galaxies have elevated central SSFR.  However, in these cases
the bright emission extends outside the nucleus.  In Fig.~\ref{fig:rose15} R band and
H$\alpha$ images are shown for these two galaxies, NGC~7074 and UGC12535.
The HII emission in NGC~7074 forms the nearly linear triple structure noticed by
Zwicky (1972).  It appears that this galaxy is seen close to edge-on, thus no
conclusion can be reached about a bar structure.  However, it is also clear
that the faint outer R band light extends well beyond the H$\alpha$ structure
and is surprisingly regular.  UGC12535 is seen almost perfectly edge-on, thus
the inner HII morphology is not discernable.

\subsection{Individual Galaxies}

\begin{figure}
\plotone{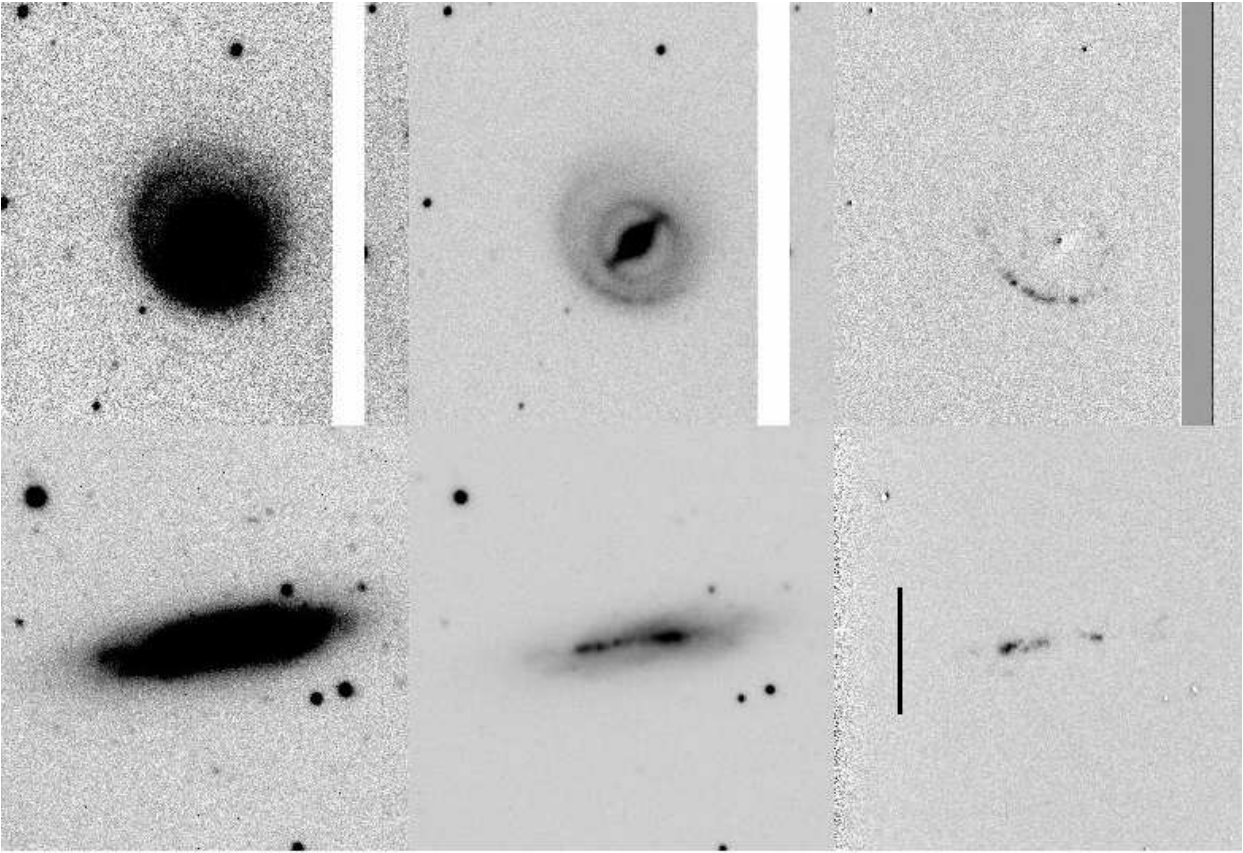}
\caption{R band and H$\alpha$ images are displayed for two galaxies with 
unusual morphologies.  CGCG~059-019 is in the upper panel and UGC12497 is in the 
lower panel.  The R band images are shown at two contrast levels in the left
and middle panels, while the H$\alpha$ images are shown in the right panels.
The black line in the lower right panel indicates an angular scale of 30\arcsc.}
\label{fig:rose16}
\end{figure}

We conclude the discussion of SFR profiles in HI-deficient spirals by briefly
investigating two individual cases of spirals whose complex H$\alpha$ morphologies 
are not adequately represented by an azimuthally averaged radial 
profile.  R band and H$\alpha$ images for CGCG~059-019=AGC170317 and for
UGC12497 are shown in Fig.~\ref{fig:rose16}.  The H$\alpha$ emission in
CGCG~059-019 is confined to an arc of HII regions along the SSE edge of this
gas-poor ($DEF$=0.65) galaxy.  The R band image reveals an inner bar and ring, 
surrounded by a single outer spiral arm.  The H$\alpha$ emission traces just
the southern edge of this arm.  The asymmetry in the R band and H$\alpha$ 
emission is reminiscent of NGC~2276 in the NGC~2300 group \citep{d97}.
In that case, the similar asymmetry seen in both the stellar optical 
light and the H$\alpha$ light led \citet{d97} to propose a 
tidal origin for the asymmetry in NGC~2276.  In the case of CGCG~059-019, 
however, the appearance of a {\it single} outer spiral arm, as well as the almost
perfect arc of HII regions, seems more consistent with a ram pressure
event, such as the single leading spiral arm seen in the ram pressure numerical
simulation of \citet[][see also Schulz \& Struck 2001]{to94}.  On the other hand, the \citet{to94} simulation 
only follows the evolution of the gaseous disk.  It appears unlikely that the
stellar disk would exhibit the same asymmetric response.  Nevertheless, in
CGCG~059-019 the R band light clearly shows the m=1 single-arm pattern.
Note that if the single arm is caused by the ram pressure
effect modeled by \citet{to94}, then the galaxy disk must be oriented nearly
parallel with respect to its motion relative to the intragroup gas.  Hence the
galaxy, according to the near circular appearance of the inner ring,
must be moving almost exactly in the plane of the sky.  
CGCG~059-019 is a member of compact group \#65 in 
the catalog of \citet{fk02}, which contains 7 galaxies and has a 
velocity dispersion of 329 \kms.  Thus it appears to be in a similar group 
environment to NGC~2276 in the NGC~2300 group.  In any case, if it can be
established that the single arm in CGCG~059-019 is indeed leading, then it
may represent a key object for distinguishing ram pressure from tidal effects
in the group environment.  Just as intensive studies of specific
galaxies in rich clusters have led to unambiguous determinations of ram
pressure effects \citep[e.g.,][]{kk99, k04, vo06}, of tidal effects 
\citep{ke96}, and of both \citep{yo08},
a further investigation of CGCG~059-019 may
provide unambiguous clues to the mechanism producing the m=1 spiral pattern
in that galaxy.

The other galaxy with exceptional morphology is UGC12497, also seen in 
Fig.~\ref{fig:rose16}.  This galaxy is a member of the central group of the
Pegasus I cluster, and has been classified as Im.  While the outer isophotes
of the R band image do indeed look like those of a single object, both the
inner R band isophotes and H$\alpha$ image are perhaps more indicative of
two discrete spirals in the process of merging.  If so, their line-of-sight
relative velocity must be low at this point, since the HI profile width given
by L07 is only 188 \kms.  The HI profile, shown in Fig.~3.18 of
L07, does have a small spike at $\sim$3740 \kms, indicative of a 
second still-discrete galaxy.

\section{Discussion}

As was mentioned in the Introduction, both KK04ab and FG08 conclusively
establish a relation between global SSFR and HI deficiency in Virgo
cluster spirals, but disagree as to whether truncation of the star-forming
disk or global quenching in SSFR is the more important driver.
FG08 argue that disk quenching is the
key process in Virgo cluster spirals while KK04ab find that truncation is
important in approximately half of their Virgo spirals.  We now
consider how the results of our study impact on the issue of disk
truncation versus disk quenching.

In this study we have found that the relation between global SSFR and atomic gas
deficiency is evident as well in the Pegasus I cluster and in other low-richness
environments.  Naturally, it is an oversimplification to compare the
``Pegasus environment'' with that of the ``Virgo environment'', since an
individual Virgo cluster galaxy may have experienced a wide range of different
environmental histories over the past $\sim$10 Gyr, depending on the specifics
of its group membership and orbital path through the greater cluster.
Neverthless, galaxies in the Pegasus I cluster and in the ``non-cluster''
sample on average have experienced an environment that has a considerably 
lower peak ram pressure (due to both lower galaxy velocity dispersion and to
lower density of the ICM) than for Virgo.  Thus it is interesting to see the
correlation between global SSFR and DEF that is found in Virgo continue in the
environments inhabited by our galaxy sample.

Regarding the issue of truncation versus starvation of disks, our results 
indicate a steady reduction in star-forming disk length with increasing DEF.
This result, which is consistent with that found by KK04b in Virgo cluster
spirals, implies that gas is being preferentially removed from the outside of
stellar disks by the processes at work in the less extreme environments
sampled in our study.  In comparing our results to that found by FG08 for
Virgo spirals, it is perhaps important that FG08 considered highly luminous
galaxies, while the typical galaxy in our sample is $\sim$1-1.5 mag less
luminous.  In addition, the spirals with truncated disks also have lower SSFR 
within the actively star-forming part of the disk than
do gas-rich spirals with normal size disks.

An important difference between our results and that found previously
in the Virgo cluster environment is the high ($\sim$50\%) fraction of spirals
with DEF$\gtrsim$0.2 that have high central SSFR.  KK04b find three Virgo
spirals with high DEF that have severely truncated H$\alpha$ disks, but high
central SFR.  All three galaxies have M$_B$ between -18 and -19, which is
typical of our sample, but substantially fainter than the luminous spirals
studied by FG08.  The three KK04b examples have higher DEF (0.79, 1.23, and
$>$1) and more severely truncated H$\alpha$ disks than in our sample, but 
otherwise appear to have similar characteristics.  However, those three
galaxies only constitute $\sim$8\% of the Virgo spirals studied by KK04b,
while we find that $\sim$50\% (although the number statistics are small) of
our spirals with DEF$\gtrsim$0.2 have high central SSFR.
The two noteworthy differences between
our galaxy sample and the Virgo sample of KK04b are that (1) our galaxies 
are typically less gas-depleted than the Virgo spirals, by about a factor of 
2-4, and (2) the environments inhabited by our sample are poor clusters and
groups, as opposed to the denser Virgo environment.  

The elevated SSFR in the centers of many partially HI-depleted spirals poses
a particularly interesting problem.
Without an ongoing HI resupply, such high
(an order of magnitude higher than in normal spirals)
central SSFRs are unsustainable, and thus indicate a transient evolutionary state.
There are a number of galaxies in Pegasus I which may in fact represent end
products of this transient state of high central SSFR.  \citet{vbr89}
found several galaxies in Pegasus, classified as early-type, but
with spectroscopic evidence of recent (NGC~7557, NGC~7611, and NGC~7617) or
still ongoing (NGC~7648) central star formation.  Optical broadband imaging 
presented in \citet{ro01} further indicates recent centralized star
formation.  NGC~7648 may be a particularly relevant example.  In addition to
the broadband data in \citet{ro01},
B, R, and H$\alpha$ imaging in L07 reveals that despite its S0
classification, NGC~7648 has highly centrally concentrated H$\alpha$ emission,
a bright off-nuclear blue knot, and ripples reminiscent of tidal interaction.
The peaked HI profile for NGC~7648 suggests that the small amount of remaining
HI is not in a cold rotating disk, thus probably also centrally concentrated.
All of these characteristics indicate that NGC~7648 may represent the last
transition stage before evolution into an S0.  The data for NGC~7563 presented
in this paper are also indicative of a pre-S0 state.  Neither HI nor H$\alpha$ 
are detected in this galaxy, but the central morphology is sufficiently 
structured that it has been classified as an Sa.  In fact, both NGC~7643 and
NGC~7563 are likely examples of the ``problem'' of galaxy morphology emphasized by
\citet{kk98}.  HI-deficient galaxies in clusters tend to be
mis-classified as earlier types than their central concentration parameters
would indicate.  Naturally, once a disk galaxy becomes completely depleted in
atomic and molecular gas, it will eventually be classified as an S0, and then
no longer be considered in samples of HI deficient galaxies, which are 
restricted to Sa and later types.  Both NGC~7648 and NGC~7563 appear to be on
the brink of the S0/a morphology.

We note other possible examples of cluster galaxies with elevated
central SFR.  In an objective prism survey of eight nearby galaxy clusters, 
\citet{mw93, mw00, mw05} find numerous spirals with centrally
concentrated H$\alpha$ emission.  In addition, \citet{ca93} and
\citet{cr97} find examples of ``early-type'' galaxies in nearby
clusters that have spectroscopic evidence for recent star formation.  Long-slit
spectroscopy \citep{ca96, ro01} and {\it HST} imaging
\citep{crd99} indicate that the recent star formation has
been concentrated in the inner $\sim$2 kpc of these galaxies.  \citet{bd86}
studied blue compact star-forming disk galaxies in the Coma
cluster.  However, these spirals have elevated SFR throughout their truncated
star-forming disks, which is distinct from our spirals that have truncated
star-forming disks but {\it lower} SFR within the active disk, except for the
very central region of elevated SFR.

In closing, we  return to the key question regarding whether ram pressure and/or
tidal gravitational effects are primarily responsible for the HI depletion
in disk galaxies in poor cluster and group environments.  Have the
data presented here informed this question in any decisive way?  We begin by 
summarizing the five principal conclusions from this study that are pertinent
to the issue.  First, the {\it global} SSFR is correlated with DEF, i.e., lowered SSFR
accompanies higher atomic gas depletion.  Second, there is a clear
correlation between the H$\alpha$-to-optical disk size ratio with DEF, i.e.,
more HI-deficient galaxies have more truncated disks.  Third, truncation
{\it and} suppression of SSFR (within the star-forming part of the disk) are
both seen in the same disks, and contribute about equally to the SSFR-DEF
correlation.  Fourth, the central SSFR is greatly enhanced in some high DEF
galaxies.  Fifth, strong bars and a ring of HII regions at the end of the bar
appear to be present in the high DEF galaxies with high central SSFR.  Finally,
these effects are observed by us in the poor cluster and group environment.

Are ram pressure stripping (RPS) and/or tidal effects consistent with these 
observational trends?  Let us first consider the case of strong RPS.  One
might expect that with strong RPS the disk will be HI-depleted in
proportion to the strength of the ram pressure.  Thus a correlation between
H$\alpha$-to-optical disk size ratio might be expected, in addition to a
correlation between global SSFR and DEF.  Thus the first two effects, and
probably the third, will result in the strong RPS case.  On the other hand,
it is not clear why the fourth and fifth effects would be observed.  Bar
structure and highly elevated central SFR are expected in the event that
a non-axisymmetric structure is triggered; strong RPS does not obviously
provide that trigger.  And the fact that we onserve the various effects in
galaxies in such low density environments is inconsistent with strong RPS.

Tidal interaction will naturally trigger non-axisymmetric disturbances, which
will readily lead to the bar structures and high central SSFR.  On the other
hand, it is not all all clear how the H$\alpha$-to-optical disk size ratio
correlation with DEF will result.  A tidal disturbance will tend to provide
non-axisymmetric disturbances whose scale is {\it internally} set (by the
location of the inner Lindblad resonance, i.e., by the internal mass
distribution and resultant rotation curve), rather than by the strength of
the tidal disturbance.  Thus the H$\alpha$ disk truncation radius will not
obviously correlate with the magnitude of the disturbance, which presumably
correlates with DEF.  Thus both strong RPS and tidal disturbances appear to
have opposite strengths and weaknesses when confronted by the observations.
The case of weaker RPS represents a potentially interesting middle ground.
\citet{ss01} find that weaker RPS leads to an atomic gas disk compression
that in turn is unstable to growth of non-axisymmetric structure, which results
in the ``annealing'' of the disk, i.e., a shrunken disk of atomic gas.  Thus
weak RPS does produce a truncation of the disk.  What is not clear, however,
is whether the degree of the truncation will correlate with the strength of
the ram pressure or whether, like in the case of tidal perturbations, one
expects the scale to be set internally, by the nature of the galaxy's
rotation curve.  Also unclear is whether the SSFR in the star-forming part of
the disk will be suppressed.  The disk annealing should in any event be 
accompanied by bar structure, gas inflow, and an elevated SSFR.  

In short, there is considerable ``degeneracy'' in behavior between RPS-induced 
and gravitationally-induced disk evolution.  Both mechanisms are capable of
redistributing the angular momentum of the gas such that gas is driven to the
galaxy centers, and both are capable of selectively removing the outer gas
supply.   Consequently, although the observational results prsented here
provide some interesting new constraints on any model for gas removal in disk
galaxies, we do not see a clear signature at this point for any single
mechanism.

\section{Conclusions}

We have obtained H$\alpha$ and R band imaging of 29 spiral galaxies in the 
spiral-rich Pegasus I cluster and of 18 spirals in non-cluster environments.
The H$\alpha$ image is used to track the SFR, while the R band roughly tracks
the total stellar light, and hence the stellar mass.  We compare the logarithmic
H$\alpha$ to R band flux ratio (i.e., the logarithm of the SFR per unit stellar
mass, or SSFR) to the logarithmic HI deficiency parameter, $DEF$, for our sample of
disk galaxies.  We find that global SSFR is well correlated with
HI deficiency, from the most HI-rich to the most HI-poor galaxies, in the
expected sense that gas-poor galaxies have lower global SSFRs.  This establishes
in lower density environments the same connection between HI deficiency and
lowered SFR that has been observed before in rich clusters such as Coma and
Virgo.  We also find that the H$\alpha$ disk size relative to the R band disk
size correlates with $DEF$, i.e., H$\alpha$ disks are truncated relative to
the optical disk size in more gas-poor spirals.  If the SFR in the nuclear 
region of the galaxy is excluded, we find that the SSFR within the 
H$\alpha$ disk is reduced in HI deficient galaxies.  Thus gas-depleted galaxies
not only have a smaller H$\alpha$ disk with increasing HI deficiency, but the
SSFR within the star-forming part of the disk is also
lower.  However, for approximately half of the gas-poor galaxies, the central
SSFR is greatly elevated relative to that in gas-rich spirals.  Furthermore,
the SSFR in the nucleus relative to the global value in
the galaxy {\it increases} with HI deficiency.  These high nuclear SSFRs in
partially HI-depleted galaxies point to a transient evolutionary state in
which gas is efficiently transported to the central region.  Whether the
condition is triggered more readily by ram pressure effects, or whether the 
origin lies with gravitational perturbation, remains to be seen.  We also note
the special case of CGCG~059-019, a one-armed spiral with a nuclear ring and
bar, in a small galaxy group.  The m=1 patern may be the direct result of
a ram pressure wind parallel to the galaxy's disk.
Finally, since global and local SFRs in spirals are so clearly coupled to
HI depletion, and since the SFR (through H$\alpha$ imaging) and kinematics of
star-forming regions can be studied at higher spatial resolution than the HI
distribution, H$\alpha$ imaging and kinematics provide a particularly
powerful probe of the mechanisms driving disk galaxy evolution.  

We wish to thank the anonymous referee for suggestions that have substantially
improved the final paper.

\clearpage
\begin{deluxetable}{ccccccc}
\tabletypesize{\scriptsize}
\tablewidth{0pc}
\tablecaption{Galaxy Properties\label{tab:one}}
\tablehead{
  \colhead{(1)}&\colhead{(2)}&\colhead{(3)}&\colhead{(4)}&\colhead{(5)}&
  \colhead{(6)}&
  \colhead{(7)} \\
  \colhead{Name}&\colhead{Alt Name}&\colhead{RA}&
  \colhead{DEC}&\colhead{Type}&
  \colhead{cz}& \colhead{Log Density}
}
\startdata
\multicolumn{7}{c}{Non-Pegasus Galaxies} \\
\\

NGC~41       & AGC~100086 & 00:12:48.0 & +22:01:24 &  5 & 5949 & -1.75 \\
UGC~144      &   ---      & 00:15:26.8 & +16:14:07 &  4 & 5620 & -1.60 \\
UGC~164      &   ---      & 00:17:23.7 & +18:05:03 &  4 & 5443 & -0.63 \\
NGC~352      & AGC~400553 & 01:02:09.2 & -04:14:44 &  3 & 5284 & -1.46 \\
UGC~1026     &   ---      & 01:27:13.0 & +13:36:08 &  9 & 4505 & -1.98 \\
IC~1721      & UGC~1187   & 01:41:24.4 & +08:31:32 &  3 & 4299 & -2.53 \\
CGCG~059-019 & AGC~170317 & 08:01:54.1 & +09:37:33 &  4 & 4793 & -1.28 \\
NGC~5417     & UGC~8943   & 14:02:13.0 & +08:02:14 &  1 & 4879 & -2.03 \\
IC~1132      & UGC~9965   & 15:40:06.7 & +20:40:50 &  5 & 4259 & -1.86 \\
NGC~5990     & UGC~10024  & 15:46:16.3 & +02:24:56 &  1 & 3839 & -2.45 \\
IC~1205      & AGC~260337 & 16:14:15.9 & +09:32:14 &  2 & 3240 & -2.54 \\
UGC~11524    &   ---      & 20:12:03.9 & +05:45:49 &  5 & 5257 & -1.42 \\
CGCG~373-007 & AGC~300178 & 20:26:14.7 & +01:06:11 &  5 & 3708 & -2.18 \\
NGC~7074     & AGC~310103 & 21:29:38.8 & +06:40:57 &  2 & 3476 & -2.36 \\
NGC~7081     & UGC~11759  & 21:31:24.1 & +02:29:29 &  3 & 3273 & -2.49 \\
CGCG~402-018 & AGC~310171 & 21:45:58.8 & +07:52:18 &  4 & 3754 & -2.32 \\
UGC~11848    &   ---      & 21:55:35.2 & +10:27:59 &  8 & 4992 & -2.59 \\
UGC~12479    &   ---      & 23:17:25.7 & -01:35:10 &  1 & 4206 & -2.53 \\
\\

\multicolumn{7}{c}{Pegasus I Cluster Galaxies} \\
\\

UGC~12304    &   ---      & 23:01:08.3 & +05:39:16 &  5 & 3470 & -0.94 \\
UGC~12361    &   ---      & 23:06:22.4 & +11:17:08 & 10 & 2992 & -1.60 \\
UGC~12370    &   ---      & 23:07:06.3 & +09:57:39 &  6 & 4892 & -1.77 \\
IC~1474      & UGC~12417  & 23:12:51.3 & +05:48:23 &  6 & 3506 & -0.20 \\
NGC~7518     & UGC~12422  & 23:13:12.7 & +06:19:18 &  1 & 3071 & -0.34 \\
NGC~7529     & UGC~12431  & 23:14:03.2 & +08:59:33 &  4 & 4538 & -1.54 \\
NGC~7537     & UGC~12442  & 23:14:34.5 & +04:29:54 &  4 & 2674 & -1.60 \\
NGC~7541     & UGC~12447  & 23:14:43.9 & +04:32:04 &  5 & 2689 & -1.49 \\
UGC~12451    &   ---      & 23:14:45.2 & +05:24:47 & 10 & 3645 & -0.94 \\
NGC~7563     & UGC~12465  & 23:15:55.9 & +13:11:46 &  1 & 4174 & -1.00 \\
UGC~12467    & NGC7562A   & 23:16:01.4 & +06:39:08 &  8 & 3507 & -0.39 \\
CGCG~406-042 & AGC~330179 & 23:17:05.5 & +07:07:22 &  5 & 3564 & -0.56 \\
NGC~7580     & UGC~12481  & 23:17:36.4 & +14:00:04 &  4 & 4434 & -0.97 \\
NGC~7593     & UGC~12483  & 23:17:57.0 & +11:20:57 &  5 & 4108 & -1.46 \\
NGC~7591     & UGC~12486  & 23:18:16.3 & +06:35:09 &  4 & 4956 & -1.49 \\
UGC~12494    & AGC~331419 & 23:18:52.6 & +06:52:38 &  7 & 4196 & -1.62 \\
UGC~12497    &   ---      & 23:19:10.8 & +07:42:13 & 10 & 3761 & -0.47 \\
IC~5309      & UGC~12498  & 23:19:11.6 & +08:06:34 &  3 & 4198 & -1.37 \\
NGC~7608     & UGC~12500  & 23:19:15.3 & +08:21:01 &  4 & 3508 & -0.17 \\
NGC~7610     & UGC~12511  & 23:19:41.4 & +10:11:06 &  6 & 3554 & -0.57 \\
NGC~7615     & AGC~330237 & 23:19:54.4 & +08:23:58 &  3 & 4473 & -1.33 \\
UGC~12522    &   ---      & 23:20:16.6 & +08:00:20 &  9 & 2812 & -1.29 \\
UGC~12535    &   ---      & 23:21:01.6 & +08:10:46 &  4 & 4214 & -1.44 \\
UGC~12544    &   ---      & 23:21:45.1 & +09:04:40 & 10 & 2859 & -1.40 \\
UGC~12553    &   ---      & 23:22:13.7 & +09:23:03 & 10 & 3573 & -0.15 \\
UGC~12562    &   ---      & 23:22:47.5 & +11:46:22 &  8 & 3836 & -0.76 \\
NGC~7643     & UGC~12563  & 23:22:50.4 & +11:59:20 &  2 & 3878 & -0.46 \\
UGC~12561    &   ---      & 23:22:58.5 & +08:59:37 &  8 & 3743 & -0.43 \\
UGC~12580    &   ---      & 23:24:33.8 & +08:36:58 &  1 & 3033 & -1.36 \\
\enddata
\end{deluxetable}

\clearpage
\begin{deluxetable}{ccccccccccc}
\tabletypesize{\scriptsize}
\tablewidth{0pc}
\tablecaption{Additional Galaxy Data\label{tab:two}}
\tablehead{
  \colhead{(1)}&\colhead{(2)}&\colhead{(3)}&\colhead{(4)}&\colhead{(5)}&
  \colhead{(6)}&
  \colhead{(7)}&\colhead{(8)}&\colhead{(9)}&\colhead{(10)}&
  \colhead{(11)} \\
  \colhead{Name}& \colhead{DEF}&
  \colhead{$\Gamma$}& \colhead{$\Gamma^*$}& \colhead{M$_H$}&
  \colhead{r$_{10}$}& \colhead{$\frac{r_{H\alpha}}{r_{10}}$}&
  \colhead{r$_{eff}$}& \colhead{$\gamma(0.1)$}& \colhead{$\Delta$}&
  \colhead{$\Gamma_{H\alpha}$} \\
  \colhead{} & \colhead{} & \colhead{} & \colhead{} & \colhead{} &
  \colhead{(\arcsc)} & \colhead{} & \colhead{(\arcsc)} & \colhead{} &
  \colhead{} & \colhead{} \\
}
\startdata
\multicolumn{11}{c}{Non-Pegasus Galaxies} \\
\\

NGC~41       & 0.01 &  -1.84 & -1.72 & -23.62 &  33. & 0.82 &  10. & -1.97 & -0.13 & -1.96 \\   
UGC~144      & 0.96 &  -3.10 & -3.08 & -22.67 &  31. & 0.43 &  12. & -2.39 &  0.71 & -2.75 \\   
UGC~164      & 0.99 &  -2.24 & -2.18 & -23.05 &  45. & 0.80 &  26. & -1.28\tablenotemark{b} &  0.96 & -2.11 \\   
NGC~352      &-0.05 &  -2.49 & -2.25 & -24.83 &  76. & 0.79 &  26. & -2.04 &  0.45 & -2.66 \\   
UGC~1026     & 0.40 &  -2.83 & -3.13 & -19.63\tablenotemark{a} &  27. & ---- &  14. & ----- & ----- & ----- \\   
IC~1721      & 0.30 &  -1.85 & -1.83 & -22.72 &  32. & 0.74 &  12. & -2.06 & -0.21 & -1.82 \\   
CGCG~059-019 & 0.78 &  -2.62 & -2.63 & -22.41\tablenotemark{a} &  29. & 0.66 &  13. & ----- & ----- & -2.72 \\   
NGC~5417     & 0.65 &  -3.88 & -3.66 & -24.63 &  52. & 0.28 &  12. & -2.68 &  1.20 & -3.29 \\   
IC~1132      &-0.18 &  -1.95 & -1.88 & -23.14 &  39. & 0.83 &  17. & -1.95 &  0.00 & -2.00 \\   
NGC~5990     & 0.21 &  -2.14 & -2.00 & -23.83\tablenotemark{a} &  53. & 0.51 &  11. & -2.45 & -0.31 & -2.42 \\   
IC~1205      &-0.42 &  -1.82 & -1.85 & -22.20 &  22. & 0.69 &   8. & -2.02 & -0.20 & -1.88 \\   
UGC~11524    &-0.52 &  -1.74 & -1.61 & -23.73 &  38. & 1.08 &  18. & -2.45 & -0.71 & -1.57 \\   
CGCG~373-007 & 0.67 &  -3.05 & -3.15 & -21.54 &  29. & 0.14 &  12. & -2.35 &  0.70 & ----- \\   
NGC~7074     & 0.52 &  -1.92 & -1.92 & -22.52 &  35. & 0.53 &  12. & -1.43\tablenotemark{b} &  0.49 & -1.99 \\   
NGC~7081     &-0.28 &  -1.79 & -1.82 & -22.21 &  39. & 0.69 &  10. & -1.86 & -0.07 & -1.71 \\   
CGCG~402-018 & 0.11 &  -2.00 & -2.07 & -21.86\tablenotemark{a} &  29. & 0.68 &  14. & -2.19 & -0.19 & -2.05 \\   
UGC~11848    &-0.05 &  -2.01 & -2.24 & -20.31\tablenotemark{a} &  30. & 1.00 &  15. & -2.13 & -0.12 & -1.87 \\   
UGC~12479    &-0.41 &  -2.22 & -2.23 & -22.38 &  36. & 0.56 &  10. & -2.01 &  0.21 & -2.26 \\   
\\

\multicolumn{11}{c}{Pegasus I Cluster Galaxies} \\
\\

UGC~12304    & 0.38 &  -2.31 & -2.28 & -22.81 &   ---- & ---- &   ---- & ----- & ----- & ----- \\   
UGC~12361    & 0.01 &  -2.15 & -2.31 & -20.98\tablenotemark{a} &   ---- & ---- &   ---- & ----- & ----- & ----- \\   
UGC~12370    &-0.04 &  -2.26 & -2.34 & -21.70 &  41. & 0.62 &  14. & -1.85 &  0.40 & -2.10 \\   
IC~1474      &-0.05 &  -1.94 & -1.93 & -22.63 &  34. & 0.93 &  14. & -2.33 & -0.39 & -2.07 \\   
NGC~7518     & 0.27 &  -2.23 & -2.16 & -23.21 &  43. & 0.61 &  22. & -1.49\tablenotemark{b} &  0.74 & -2.18 \\   
NGC~7529     &-0.08 &  -1.99 & -2.03 & -22.16 &  29. & 0.72 &  11. & ----- & ----- & -2.19 \\   
NGC~7537     &-0.31 &  -1.76 & -1.77 & -22.36 &  57. & 1.00 &  14. & -1.93 & -0.17 & -1.78\\    
NGC~7541     &-0.10 &  -1.92 & -1.76 & -24.06\tablenotemark{a} &   ---- & ---- &   ---- & ----- & ----- & ----- \\   
UGC~12451    & 0.24 &  -1.98 & -2.30 & -19.43 &  50. & 0.88 &  36. & -2.19 & -0.21 & ----- \\   
NGC~7563     & 1.82 &  -5.40 & -5.22 & -24.27 &  53. & 0.00 &  10. & ----- & ----- & ----- \\   
UGC~12467    & 0.32 &  -2.14 & -2.18 & -22.08\tablenotemark{a} &  59. & 0.64 &  26. & -2.20 & -0.06 & -2.13 \\   
CGCG~406-042 & 0.41 &  -2.38 & -2.59 & -20.51 &  29. & 0.77 &  16. & -2.85 & -0.47 & -2.17 \\   
NGC~7580     &-0.29 &  -1.76 & -1.74 & -22.70 &  26. & 0.64 &   8. & -1.23 &  0.53 & -2.15 \\   
NGC~7593     & 0.07 &  -1.99 & -1.98 & -22.64 &  34. & 0.61 &  14. & -1.79 &  0.20 & -1.96 \\   
NGC~7591     &-0.25 &  -1.96 & -1.77 & -24.32 &  62. & 1.04 &  14. & -1.52 &  0.44 & -1.73 \\   
UGC~12494    & 0.06 &  -1.90 & -2.01 & -21.44 &  43. & 0.88 &  20. & -1.67 &  0.23 & -1.85 \\   
UGC~12497    &-0.04 &  -2.22 & -2.25 & -22.19\tablenotemark{a} &  42. & 0.46 &  20. & -1.84 &  0.38 & -2.69 \\   
IC~5309      & 0.29 &  -2.00 & -1.95 & -22.95 &  56. & 0.54 &  19. & -1.83 &  0.17 & -2.07 \\   
NGC~7608     & 0.48 &  -2.07 & -2.05 & -22.56 &  47. & 0.87 &  20. & -1.69\tablenotemark{b} &  0.38 & -2.04 \\   
NGC~7610     &-0.15 &  -1.30 & -1.32 & -22.29 &  55. & 1.00 &  22. & -1.64 & -0.34 & -1.39 \\   
NGC~7615     & 0.85 &  -2.52 & -2.53 & -22.45 &  29. & 0.79 &  11. & ----- & ----- & -2.46 \\   
UGC~12522    & 0.19 &  -2.24 & -2.43 & -20.69 &  35. & 0.30 &  20. & -2.39 & -0.15 & -2.28 \\   
UGC~12535    & 0.51 &  -2.11 & -2.15 & -22.07 &  40. & 0.68 &  16. & -1.64\tablenotemark{b} &  0.47 & -2.13 \\   
UGC~12544    &-0.06 &  -2.11 & -2.27 & -20.99\tablenotemark{a} &  38. & 0.65 &  20. & ----- & ----- & -2.25 \\   
UGC~12553    & 0.16 &  -2.80 & -2.80 & -20.99\tablenotemark{a} &  26. & ---- &  15. & ----- & ----- & ----- \\   
UGC~12562    & 0.16 &  -2.05 & -2.41 & -19.03 &  37. & 0.67 &  15. & -2.34 & -0.29 & -2.05 \\   
NGC~7643     & 0.65 &  -2.37 & -2.27 & -23.44 &  46. & 0.57 &  20. & -1.63\tablenotemark{b} &  0.74 & -2.33 \\   
UGC~12561    & 0.19 &  -2.98 & -3.07 & -21.60\tablenotemark{a} &  43. & 0.12 &  20. & -2.32 &  0.66 & ----- \\   
UGC~12580    & 1.11 &  -3.69 & -3.74 & -22.05 &  37. & 0.05 &   9. & -2.39 &  1.36 & ----- \\   
\enddata
\tablenotetext{a} {No H band data is available from Hyperleda; M$_H$ was 
determined from total absolute B magnitude and an assumed $B-H$ color of
-2.52.}
\tablenotetext{b} {Galaxies with high $DEF$ and high central SSFR ($\gamma(0.1)$)}

\end{deluxetable}


\clearpage
\begin{thebibliography}{}
\bibitem[Abadi et al. (1999)]{ab99} Abadi, M. G., Moore, B., \& Bower, R. G. 
  1999, \mnras, 308, 947
\bibitem[Boselli \& Gavazzi (2006)]{bg06} Boselli, A., \& Gavazzi, G. 2006, 
  \pasp, 118, 517
\bibitem[Bothun \& Dressler (1986)]{bd86} Bothun, G. D., \& Dressler, A. 1986,
  \apj, 301, 57
\bibitem[Bravo-Alfaro et al. (2001)]{ba01} Bravo-Alfaro, H., Cayatte,
  V., van Gorkom, J. H., \& Balkowski, C. 2001, \aap, 379, 347
\bibitem[Butcher \& Oemler (1978)]{bo78} Butcher, H., \& Oemler, A.,
  Jr. 1978, \apj, 219, 18
  \bibitem[Butcher \& Oemler (1984)]{bo84} Butcher, H., \& Oemler, A.,
    Jr. 1984, \apj, 285, 426
\bibitem[Byrd \& Valtonen (1990)]{bv90} Byrd, G, \& Valtonen, M. 1990, \apj,
  350, 89
\bibitem[Caldwell et al. (1993)]{ca93} Caldwell, C., Rose, J. A., Sharples, R.
  M., Ellis, R. S., \& Bower, R. G. 1993, \aj, 106, 473
\bibitem[Caldwell et al. (1996)]{ca96} Caldwell, C., Rose, J. A., Franx, M., \&
  Leonardi, A. J. 1996, \aj, 111, 78
\bibitem[Caldwell \& Rose (1997)]{cr97} Caldwell, N., \& Rose, J. A. 1997, \aj, 
  113, 492
\bibitem[Caldwell, Rose, \& Dendy (2003)]{crd99} Caldwell, N., Rose, J. A., \& Dendy,
  K. 1999, \aj, 117, 140
\bibitem[Cayatte et al. (1994)]{ca94} Cayatte, V., Kotanyi, C.,
  Balkowski, C., \& van Gorkom, J. H. 1994, \aj, 107, 1003
\bibitem[Chung et al. (2007)]{c07} Chung, A., van Gorkom, J. H., Kenney, J. D.
  P., \& Vollmer, B. 2007, \apj, 659, L115
\bibitem[Clemens et al. (2004)]{cl04} Clemens, J. C., Crain, J. A., \& Anderson,
  R. 2004, SPIE, 5492, 331
\bibitem[Cortes et al (2008)]{ckh08} Cortes, J. R., Kenney, J. D. P., \& Hardy,
  E. 2008, \apj, 683, 78
\bibitem[Courteau (1996)]{co96} Courteau, S. 1996, \apjs, 103, 363
\bibitem[Cowie \& Songaila (1977)]{cs77} Cowie, L. L., \& Songaila, A. 1977,
  Nature, 266, 501
\bibitem[Crowl \& Kenney (2006)] {ck06} Crowl, H. H., \& Kenney, J. D. P. 2006,
  \apj, 649, L75
\bibitem[Davis et al. (1997)]{d97} Davis, D. S., Keel, W. C.,
  Mulchaey, J. S., \& Henning, P. A. 1997, \aj, 114, 2
\bibitem[Dressler (1980)]{d80} Dressler, A. 1980, \apj, 236, 351
\bibitem[Dressler \& Gunn (1983)]{dg83} Dressler, A., \& Gunn,
  J. E. 1983, \apj, 270, 7
\bibitem[Dressler et al. (1999)]{d99} Dressler, A., Smail, I.,
  Poggianti, B. M., Butcher, H., Couch,
  W. J., Ellis, R. S., Oemler, A. Jr. 1999, \apj, 122, 51
\bibitem[Falco et al. (1999)]{fa99} Falco, E. E., Kurtz, M. J., Geller, M. J.,
  Huchra, J. P., Peters, J., Berlind, P., Mink, D. J., Tokarz, S. P., \&
  Elwell, B. 1999, \pasp, 111, 438
\bibitem[Focardi \& Kelm (2002)]{fk02} Focardi, P., \& Kelm, B. 2002, \aap,
  391, 35
\bibitem[Fumagalli \& Gavazzi (2008)]{fg08} Fumagalli, M., \& Gavazzi, G. 2008,
  \aap, 490, 571
\bibitem[Gavazzi (1987)]{ga87} Gavazzi, G. 1987, \apj, 320, 96
\bibitem[Gavazzi et al. (1995)]{g95} Gavazzi, G., Contursi, A., Carrasco,
  L., Boselli, A., Kennicutt, R., Scodeggio, M., Jaffe, W. 1995, \aap,
    304, 325
\bibitem[Gavazzi et al. (2006)]{ga06} Gavazzi, G., O'Neil, K., Boselli, A., \& 
  van Driel, W. 2006, \aap, 449, 929
\bibitem[Giovanelli \& Haynes (1985)]{gh85} Giovanelli, R., \& Haynes,
  M. P. 1985, \apj, 292, 404
\bibitem[Gomez et al. (2003)]{gom03} Gomez, P. L. et al. 2003, \apj, 584, 210
\bibitem[Goto et al. (2003)]{got03} Goto, T., Yamauchi, C., Fujita, Y., 
  Okamura, S., Sekiguchi, M., Smail, I., Bernardi, M., \& Gomez, P. L. 2003,
  \mnras, 346, 601
\bibitem[Gunn \& Gott (1972)]{gg72} Gunn, J. E., \& Gott, J. R. 1972,
   \apj, 176, 1
\bibitem[Haynes \& Giovanelli (1986)]{hg86} Haynes, M. P., \& Giovanelli, R.
  1986, \apj, 306, L55
\bibitem[Hester (2006)]{he06} Hester, J. A. 2006, \apj, 647, 910
\bibitem[Kapferer et al. (2008)]{ka08} Kapferer, W., Kronberger, T., Ferrari, 
  C., Riser, T., \& Schindler, S. 2008, \mnras, 389, 1405
\bibitem[Kantharia et al. (2005)]{ka05} Kantharia, N. G., Ananthakrishnan, S.,
  Nityananda, R., \& Hota, A. 2005, \aap, 435, 483
\bibitem[Kawata \& Mulchaey (2008)]{km08} Kawata, D., \& Mulchaey, J. S. 2008,
  \apj, 672, 103
\bibitem[Kenney et al. (1996)]{ke96} Kenney, J. D. P., \& Koopmann, R. A.,
  Rubin, V. C., \& Young, J. S. 1996, \aj, 111, 152
\bibitem[Kenney \& Koopman (1999)]{kk99} Kenney, J. D. P., \& Koopmann, R. A. 
  1999, \aj, 117, 181
\bibitem[Kenney et al. (2004)]{k04} Kenney, J. D. P., van Gorkom, J. H.,
  Vollmer, B. 2004, \aj, 127, 3361
\bibitem[Koopmann \& Kenney (1998)]{kk98} Koopmann, R. A., \& Kenney, J. D. P.
  1998, \apj, 497, L75
\bibitem[Koopmann \& Kenney (2004a)]{kk04a} Koopmann, R. A., \& Kenney,
  J. D. P. 2004, \apj, 613, 851
\bibitem[Koopmann \& Kenney (2004b)]{kk04b} Koopmann, R. A., \& Kenney,
  J. D. P. 2004, \apj, 613, 866
\bibitem[Kronberger et al. (2008a)]{kr08a} Kronberger, T., Kapferer, W., 
  Ferrari, C., Unterguggenberger, S., \& Schindler, S. 2008a, \aap, 481, 337
\bibitem[Kronberger et al. (2008b)]{kr08b} Kronberger, T., Kapferer, W., 
  Unterguggenberger, S., Schindler, S., \& Ziegler, B. L. 2008b, \aap, 483, 783
\bibitem[Landolt (1992)]{la92} Landolt, A. U. 1992, \aj, 104, 372
\bibitem[Larson, Tinsley, \& Caldwell (1980)]{ltc80} Larson, R. B., Tinsley, B.
  M., \& Caldwell, C. N. 1980, \apj, 237, 692
\bibitem[Lavery \& Henry (1988)]{lh88} Lavery, R. J., \& Henry, J. P. 1988, 
  \apj, 330, 596
\bibitem[Levy et al. (2007)]{ll07} Levy, L., Rose, J. A., van Gorkom, J. H., 
  \& Chaboyer, B. 2007, \aj, 133, 1104
\bibitem[Marzke et al (1994)]{ma94}  Marzke, R.O., Huchra, J.P., \& Geller M.J. 
  1994, \apj, 428, 43
\bibitem[Mihos \& Hernquist (1994)]{mh94} Mihos, J. C., \& Hernquist, L. 1994,
   \apj, 425, L13
\bibitem[Mihos \& Hernquist (1996)]{mh96} Mihos, J. C., \& Hernquist, L. 1996,
   \apj, 464, 641
\bibitem[Moore et al. (1996)]{mo96} Moore, B., Katz, N., Lake, G.,
  Dressler, A., \& Oemler, A. 1996, Nature, 379, 613
\bibitem[Moore et al. (1998)]{mo98} Moore, B., Lake, G., \& Katz, N. 1998, \apj,
  495, 139
\bibitem[Moss \& Whittle (1993)]{mw93} Moss, C., \& Whittle, M. 1993, \apj,
  407, L17
\bibitem[Moss \& Whittle (2000)]{mw00} Moss, C., \& Whittle, M. 2000, \mnras,
  317, 667
\bibitem[Moss \& Whittle (2005)]{mw05} Moss, C., \& Whittle, M. 2005, \mnras,
  357, 1337
\bibitem[Mulchaey et al. (1993)]{mu93} Mulchaey, J. S., Davis, D. S.,
  Mushotzky, R. F., \& Burstein, D. 1993, \apj, 404, L9
\bibitem[Nulsen (1982)]{n82} Nulsen, P. E. J.1982, \mnras, 198, 1007
\bibitem[Oemler (1974)]{o74} Oemler, A. 1974, \apj, 194, 1
\bibitem[Omar \& Dwarakanath (2005)]{od05} Omar, A., \& Dwarakanath, K. S. 
  2005, J. Astrophys. Astr., 26, 71
\bibitem[Rasmussen, Ponman, \& Mulchaey (2006)]{rpm06} Rasmussen, J., Ponman, 
  T. J., \& Mulchaey, J. S. 2006, \mnras, 370, 453
\bibitem[Roediger \& Hensler (2005)]{rh05} Roediger, E., \& Hensler, G. 2005,
  \aap, 433, 875
\bibitem[Rose et al. (2001)]{ro01} Rose. J. A., Gaba, A. E., Caldwell, N. \&
  Chaboyer, B. 2001, \aj, 121, 793
\bibitem[Sanchis et al (2004)]{sa04} Sanchis, T., Mamon, G. A., Salvador-Sole,
  E., \& Solanes, J. M. 2004, \aap, 418, 393
\bibitem[Schechter (1976)]{sc76} Schechter, P. 1976, \apj, 203, 297
\bibitem[Schulz \& Struck (2001)]{ss01} Schulz, S., \& Struck, C. 2001, \mnras,
  328, 185
\bibitem[Schwarz et al. (2004)]{sc04} Schwarz, H. E., Ashe, M. C., Boccas, M.,
  Bonati, M., Delgado, F., Gavez, R., Martinez, M., Schurter, P., Schmidt, R.,
  Tighe, R., Walker, A. R. 2004, SPIE, 5492, 564
\bibitem[Sengupta \& Balasubramyanyam (2006)]{sb06} Sengupta, C. \&
  Balasubramanyam, R. 2006, \mnras, 369, 360
\bibitem[Sengupta et al. (2007)]{se07} Sengupta, C., Balasubramanyam, R., \&
  Dwarakanath, K. S. 2007, \mnras, 378, 137
\bibitem[Solanes et al. (1996)]{so96} Solanes, J. M., Giovanelli, R., \& Haynes,
  M. P. 1996, \apj, 461, 609
\bibitem[Solanes et al. (2001)]{s01} Solanes, J. M., Manrique, A.,
   Garcia-Gomez, C., Gonzales-Casado,G., Giovanelli, R., \& Haynes,
   M. P.  2001, \apj, 548, 97
\bibitem[Solanes et al. (2002)]{so02} Solanes, J. M., Sanchis, T., 
  Salvador-Sole, E., Giovanelli, R., \& Haynes, M. P. 2002, \aj, 124, 2440
\bibitem[Springob et al. (2005a)]{sp05a} Springob, C. M., Haynes, M. P.,
  Giovanelli, R., \& Kent, B. R. 2005a, \apjs, 160, 149
\bibitem[Springob et al. (2005b)]{sp05b} Springob, C. M., Haynes, M. P., \&
  Giovanelli, R. 2005b, \apj, 621, 215
\bibitem[Struck \& Brown (2004)]{sb04} Struck, C., \& Brown, J. R. 2004, in
  IAUS\#217, Recycling Intergalactic and Interstellar Matter, ed. P.-A. Duc,
  J. Braine, \& E. Brinks (San Francisco: Astron. Soc. Pac.), p. 466
\bibitem[Thomas et al (2008)]{th08} Thomas, C. F., Moss, C., James, P. A.,
  Bennett, S. M., Aragon-Salamanca, A., \& Whittle, M. 2008, \aap, 486, 755
\bibitem[Toomre \& Toomre (1972)]{tt72} Toomre, A., \& Toomre, J. 1972, \apj,
  178, 623
\bibitem[Tosa (1994)]{to94} Tosa, M. 1994, \apj, 426, L81
\bibitem[van Dokkum et al. (2000)]{vd00} van Dokkum, P. G., Franx, M., 
  Fabricant, D., Illingworth, G. D., \& Kelson, D. D. 2000, \apj, 541, 95
\bibitem[van Gorkom (2004)]{vg04} van Gorkom, J. H. 2004, in Carnegie
  Observatory Astrophysics Series, Vol 3: Clusters of Galaxies, Probes
  of Cosmology and Galaxy Evolution, ed. J.S. Mulchaey, A. Dressler,
  and A. Oemler (Cambridge: Cambridge University Press), p306
\bibitem[Verheijen (2004)]{v04} Verheijen, M. A. W. 2004, in Outskirts
  of Galaxy Clusters: intense Life in the Suburbs, ed. A. Diaferio
    (IAU Colloquium \#195), p394
\bibitem[Vigroux, Boulade, \& Rose (1989)]{vbr89} Vigroux, L., Boulade, O., \&
  Rose, J. A. 1989, \aj, 98, 2044
\bibitem[Vollmer et al. (2001)]{vo01} Vollmer, B., Cayatte, V., Balkowski, C.,
  \& Duschl, W. J. 2001, \apj, 561, 708
\bibitem[Vollmer et al. (2004)]{vo04} Vollmer, B., Beck, R., Kenney, J. D. P.,
  \& van Gorkom, J. H. 2004, \aj, 127, 3375
\bibitem[Vollmer et al. (2006)]{vo06} Vollmer, B., Soida, M., Otmianowska-Mazur,
  K., Kenney, J. D. P., van Gorkom, J. H., \& Beck, R. 2006, \aap, 453, 883
\bibitem[Yoshida et al. (2008)]{yo08} Yoshida, M., Yagi, M., Komiyama, Y.,
  Furusawa, H., Kashikawa, N., Koyama, Y., Yamanoi, H., Hattori, T., \&
  Okamura, S. 2008, \apj, 688, 918
\end{thebibliography}
\end{document}